\documentclass[twocolumn]{autart}

\usepackage[T1]{fontenc}
\usepackage[utf8]{inputenc}
\usepackage{lmodern}
\usepackage{amsmath,amssymb}
\usepackage{graphicx}
\usepackage{subcaption}
\usepackage{wrapfig}
\usepackage{siunitx}
\usepackage{booktabs}
\usepackage{mathtools}
\usepackage{enumitem}
\usepackage{placeins}
\usepackage{latexml}
\newenvironment{rmk}[1][]{\par\smallskip\noindent\textbf{Remark%
  \ifx&#1&\else~(#1)\fi.}\enspace\itshape}{\par\smallskip}

\graphicspath{{figures/}{./}}

\newsavebox{\bioPhoto}
\newlength{\bioTopT}
\newcommand{\biographyentry}[3]{%
  \iflatexml
    \par\bigskip
    \noindent\textbf{#2}\ #3%
    \par\bigskip
  \else
  \par\bigskip
  \sbox{\bioPhoto}{%
    \begin{minipage}[b][1.25in][c]{1.0in}%
      \centering
      \includegraphics[width=1.0in,height=1.25in,keepaspectratio]{#1}%
    \end{minipage}%
  }%
  \settoheight{\bioTopT}{\mbox{T}}%
  \hangindent=1.14in
  \hangafter=-8
  \noindent\makebox[0pt][l]{%
    \hspace{-\hangindent}%
    \raisebox{\bioTopT}[0pt][0pt]{%
      \raisebox{-\ht\bioPhoto}[0pt][0pt]{\usebox{\bioPhoto}}%
    }%
  }%
  \noindent\textbf{#2}\ #3%
  \par\bigskip
  \fi}

\newcommand{\paperfrontnote}{%
Corresponding author: J.~Gillberg
(\texttt{jonas@zyqe.se}).
Preliminary results appeared in \cite{gillberg06} and the first author's
doctoral thesis \cite{gillberg_thesis}; the present paper provides a self-contained
treatment with a corrected asymptotic formulation, the piecewise centering
correction $\alpha_b$, closed-form $K(b)$, and a formally checked local
influence-curve minimax result absent from the thesis.}

\newcommand{\robustalgorithmref}{\iflatexml 1\else\ref{alg:robust-est}\fi}

\begin{document}

\begin{frontmatter}

\runtitle{Minimax-Optimal Robust CT Identification}

\title{Minimax-Optimal Robust Identification of Continuous-Time Systems:\\ Handling Narrow-Band Disturbances in the Frequency Domain}

\iflatexml
\author{Jonas Gillberg}
\ead{jonas@zyqe.se}
\author{Fredrik Gustafsson}
\ead{fredrik@isy.liu.se}
\else
\author[zyqe]{Jonas Gillberg\thanksref{footnoteinfo}}
\ead{jonas@zyqe.se}
\author[liu]{Fredrik Gustafsson}
\ead{fredrik@isy.liu.se}

\address[zyqe]{ZyQE, Stockholm, Sweden}
\address[liu]{Department of Electrical Engineering,
  Link\"{o}ping University, SE-581\,83 Link\"{o}ping, Sweden}
\fi


\begin{abstract}
The high-frequency spectral roll-off of continuous-time ARMA (CARMA)
models can magnify the effect of narrow-band disturbances when aliasing is
weak, making standard maximum-likelihood Whittle estimation sensitive to
affected ordinates.  We show that a logarithmic transformation
$r_k = \log\rho_k$ converts the Whittle scale problem into a Gumbel
location problem, connecting robust spectral estimation to the
classical minimax theory of Huber and Rieder.  A piecewise centering
correction --- closed-form for small $b$, implicit closed-form for the
practitioner range --- preserves Fisher consistency for any clipping
level without numerical optimisation.  Clipping the Gumbel score
symmetrically and transforming back yields a two-sided clipped Gumbel
score (the standard Rieder--Hampel bounded-influence form) whose
normalised influence curve is proved locally asymptotically minimax
under shrinking gross-error contamination.  The efficiency loss is
quantified by a single scalar $K(b)$: at $b = 1.5$, only $16\%$
nominal asymptotic variance overhead.  In the stated AR(2) Monte~Carlo
design, the sample-size trends are compatible with the asymptotic rate,
and at $b=1.5$ the bias reductions are about $40\%$, $90\%$, and $96\%$
for $a_1$, $a_2$, and $\lambda$, respectively.
\end{abstract}

\iflatexml\else
\thanks[footnoteinfo]{\paperfrontnote}
\fi

\end{frontmatter}

\iflatexml
\noindent\emph{Affiliations:} ZyQE, Stockholm, Sweden (J.~Gillberg);
Department of Electrical Engineering, Link\"{o}ping University,
SE-581\,83 Link\"{o}ping, Sweden (F.~Gustafsson).\par\smallskip
\noindent\emph{\paperfrontnote}\par\medskip
\fi

\section{Introduction}

\begin{figure}[t]
  \centering
  \includegraphics[width=\columnwidth]{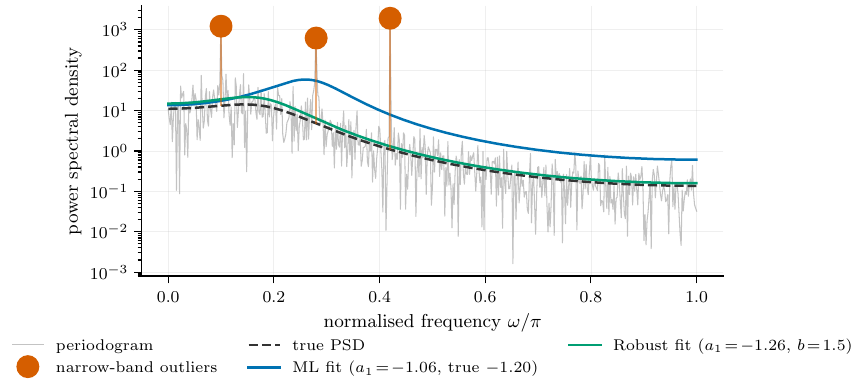}
  \caption{The outlier problem and its solution.  Log-scale periodogram
    (grey) of an AR(2) signal ($a_1=-1.2$, $a_2=0.5$) corrupted by
    three sinusoidal disturbances (orange circles; drop lines show
    the outlier magnitude relative to the true PSD).  The ML fit
    (blue) is dragged upward across the entire frequency range by
    just three outlying ordinates.  The robust Huber fit at $b=1.5$
    (green) tracks the true PSD (dashed) closely despite the
    contamination --- the core proposition of this paper.}
  \label{fig:outliersproblem}
\end{figure}

Narrow-band frequency-domain disturbances appear in several engineering
applications.  In resonance-based identification and condition monitoring,
physically informative resonant modes can coexist with harmonics generated by
rotating components, periodic actuation, or sensor artefacts
\cite{garnier08,pintelon12}.  The resonant structure then belongs to the
dynamics of interest, whereas a small number of persistent tones at
frequencies $\omega_j$ act as localised spectral contamination.  Because
these tones persist over the full observation window, frequency-domain
estimators are attractive yet sensitive to outlying periodogram ordinates
(see Fig.~\ref{fig:outliersproblem}).

\begin{figure}[t]
  \centering
  \includegraphics[width=\columnwidth]{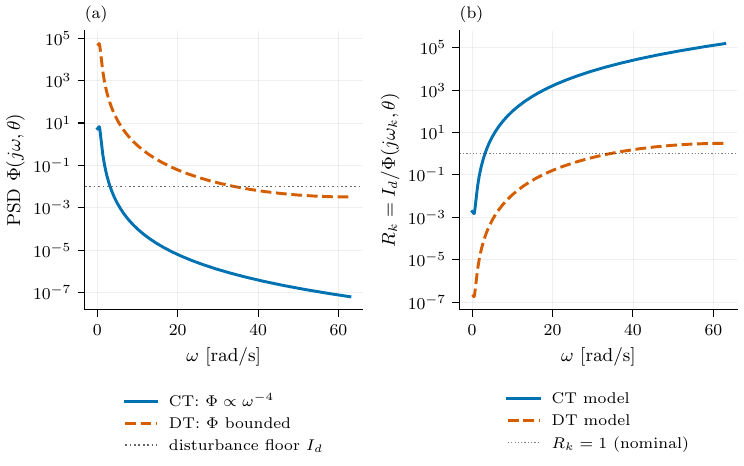}
  \caption{Illustrative continuous-spectrum vulnerability mechanism
    (not a sampled-data likelihood or an identification experiment):
    CARMA(2,\,0) with the same poles as the
    AR(2) in Section~\ref{sec:example}, sampled at $\Delta t=0.05$\,s).
    (a)~The CT spectral density (solid) rolls off as $\omega^{-4}$,
    while the equivalent discrete-time PSD (dashed) stays bounded
    above Nyquist.  A flat disturbance floor $I_d$ (dotted) is shown
    for reference.
    (b)~The normalised periodogram ratio
    $\rho_k=I_d/\Phi(j\omega_k,\theta)$ reaches $10^5$ for the CT model
    near the Nyquist frequency, versus $\mathcal{O}(1)$ for the DT
    model --- a difference of four to five orders of magnitude in this
    deliberately weak-aliasing comparison.  The exact sampled spectrum used
    for estimation is defined in \eqref{eq:sampled-spectrum}.}
  \label{fig:ct_vulnerability}
\end{figure}

While the robust Whittle criterion we develop below applies equally to
discrete-time models, the motivation is especially strong for
continuous-time systems, whose parameters (pole locations, damping
ratios, natural frequencies) have direct physical meaning and are
invariant under changes in sampling rate \cite{garnier08,young11},
unlike DT model coefficients which depend on~$\Delta t$.
The continuous-time spectral density of a CARMA($n$,\,$m$)
model rolls off as $\omega^{-2(n-m)}$ at high frequencies
(since $|B(j\omega)/A(j\omega)|^2 \propto \omega^{2(m-n)}$ as
$\omega\to\infty$), so the
normalised periodogram ratio
$\rho_k(\theta)=I_{N_s}(\omega_k)/\Phi(j\omega_k,\theta)$ can be amplified
at high frequencies if the continuous spectrum is used in a regime with
negligible aliasing.  A narrow-band disturbance can then produce ordinates
far above the nominal model and exert high leverage on the ML Whittle
criterion (Fig.~\ref{fig:ct_vulnerability}).  For sampled data, the estimator
below instead uses the exact alias-aware sampled spectrum.  These mechanisms
motivate a bounded-influence alternative without presuming that every CT data
set has the same vulnerability.

Notch filtering or periodogram trimming is the standard engineering
alternative.  The robust estimator instead retains all periodogram
ordinates with down-weighted, rather than discarded, influence.
Deterministic trimming preserves the nominal model on retained bins but
requires disturbance locations and sacrifices information; data-dependent
trimming also introduces a selection step that must be accounted for in
inference (Section~\ref{sec:discussion}).

\begin{figure}[t]
  \centering
  \includegraphics[width=\columnwidth]{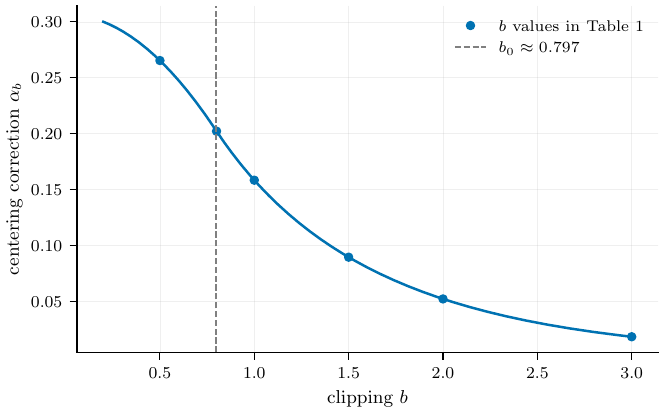}
  \caption{Centering correction $\alpha_b$ over $b\in[0.2,3]$.
    Closed form $\alpha_b = 1-b-\log\bigl((1-e^{-2b})/b\bigr)$ on
    $b<b_0\approx 0.797$ (vertical dashed line) and the implicit
    \eqref{eq:abrelation} on $b\geq b_0$; the two branches agree at
    $b=b_0$ where $\alpha_{b_0}=1-b_0\approx 0.203$.
    Dots mark the $b$ values tabulated in Table~\ref{tab:kappa}.
    The correction is exponentially small for large $b$:
    $\alpha_b < 0.06$ for $b \geq 2$ and $\alpha_b\to 0$ as $b\to\infty$.}
  \label{fig:acurve}
\end{figure}

This paper presents a robust frequency-domain CARMA
identification framework extending \cite{gillberg06,gillberg08}.  The main contributions
are:
\begin{itemize}
  \item The Gumbel location formulation: exploiting the
        exponential-scale/Gumbel-location duality
        \cite{thall79,rieder94} in the Whittle setting, the
        logarithmic transformation $r_k = \log\rho_k$ converts the
        spectral scale problem into a location problem, connecting
        robust spectral estimation to the minimax theory and
        explaining why the locally optimal score is piecewise linear
        in~$\rho$;
  \item A piecewise closed-form centering correction $\alpha_b$
        (explicit on $0<b<b_0$, implicit on $b\geq b_0$) that
        preserves Fisher consistency for every clipping level~$b$
        without numerical optimisation
        (Fig.~\ref{fig:acurve});
  \item Local minimax optimality: after the standard influence-curve
        normalisation, the two-sided clipped Gumbel score is the exact
        solution of the local asymptotic minimax problem under shrinking
        gross-error contamination
        (Theorem~\ref{thm:opt-huber});
  \item Conditional asymptotic theory: $\sqrt{N}$ normality with a
        closed-form efficiency factor $K(b)$ under the stated fixed-sampling,
        Gaussian and identifiability conditions
        (Theorem~\ref{thm:asympvar}, Table~\ref{tab:kappa}).
\end{itemize}

\paragraph*{Related work}
\textit{Frequency-domain PEM.}
The Whittle criterion \cite{whittle61} is the canonical frequency-domain
prediction error (PEM) criterion \cite{ljung99,soderstrom89} for spectral
models; for continuous-time systems see \cite{pintelon01}.  The asymptotic
Exp(1) distribution of normalised periodogram ordinates follows from
classical DFT results \cite{brillinger81,hannan70,dzhaparidze70,kay88,stoica05}.

\textit{Robust M-estimation.}
Huber's robust score and the minimax M-estimator framework originate in
\cite{huber81}; the influence function was introduced by Hampel et
al.\ \cite{hampel86}.  For modern treatments see
\cite{maronna06,maronna19,zoubir18}.  Thall~\cite{thall79} showed that
robust scale estimation for the exponential distribution reduces to
robust Gumbel location estimation via the log-transform.  The general
theory of optimally robust influence curves for smooth parametric
models (including asymmetric distributions) is developed in
Rieder~\cite{rieder94}, Ch.~5.  Kohl~\cite{kohl05},
Secs.~1.2--1.3, gives the infinitesimal-contamination influence-curve
risk and clipping--radius relation, while Sec.~5.2.1 specialises the
construction to the exponential-scale/Gumbel-location case; an
implementation is available in the R package \texttt{RobExtremes}
\cite{robextremes}.  We provide an explicit closed-form centering
correction for this setting and apply the framework to spectral
estimation.

\textit{CT identification from sampled data.}
Direct methods for identifying continuous-time models from sampled
observations are developed in \cite{garnier03,garnier08,pintelon12}.
At a fixed sampling interval, Whittle estimation for sampled continuous-time
state-space models is formulated using the exact sampled-process spectrum;
its consistency and asymptotic normality require explicit stability,
identifiability, moment and smoothness conditions \cite{fasenMayer22}.
High-frequency sampling with $\Delta t\to0$ is a distinct asymptotic regime
\cite{brockwell12}.
Time-domain instrumental-variable methods such as SRIVC
\cite{young11,garnier08} offer an alternative approach.  The
frequency-domain method developed here complements these: it handles
narrow-band spectral disturbances directly within a specified sampled CT
spectral model.

\textit{Robust identification.}
Bootstrap validity of the periodogram CLT is established
in \cite{dahlhaus96}.  Robust estimation of the autocovariance function
via clipped periodograms is developed in \cite{levy11}.
Taniguchi and Kakizawa~\cite{taniguchi00} develop a general
asymptotic theory of statistical inference for stochastic processes
in the spectral domain, including robust and higher-order methods.
Set-membership identification \cite{milanese91}
offers a non-probabilistic alternative but does not exploit the
Exp(1) residual structure.  A comprehensive frequency-domain
identification reference is \cite{pintelon12}.

Section~\ref{sec:problem} formalises the setting;
Section~\ref{sec:robust} develops the estimator and its asymptotic
theory; Section~\ref{sec:optimality} establishes local minimax optimality;
Section~\ref{sec:tuning} discusses tuning and provides a practitioner
recipe; Section~\ref{sec:example} presents simulations.
Appendix~\ref{app:lemmas} states the regularity conditions;
Appendices~\ref{app:proof-opt-huber}--\ref{app:kappa} contain
the proofs and explicit formulae.\footnote{The reference implementation,
frozen main-result bundles, and exploratory extension scripts are available at
\texttt{https://github.com/jonas-zyqe/robust-carma-revisited}.  The claims in
this article rely only on the evidence identified explicitly in the main text.}

\section{Problem Formulation}\label{sec:problem}

The setting is continuous-time ARMA (CARMA) identification in the
frequency domain, in the presence of narrow-band output disturbances.
Throughout, the nominal process $y(t)$ is assumed zero-mean, Gaussian and
second-order stationary, with rational spectral density
$\Phi(j\omega,\theta_0)$.  We assume $\Phi$ is bounded away from
zero and infinity on $[\omega_{\min},\pi/\Delta t]$ for some
$\omega_{\min}>0$, with DC excluded and no poles on the imaginary
axis.  Processes with a non-zero mean or deterministic inputs are
out of scope.
The observed output satisfies
\begin{equation}
  Y = H\,E + D,
\end{equation}
where $H$ is the parametric system, $E$ denotes nominal white noise,
and $D$ represents additive narrow-band disturbances (narrow spectral
spikes).  The continuous-time parameterisation is
\begin{align}
  y(t)        &= H(s,\theta)\,e(t), \qquad
  H(s,\theta)  = \frac{B(s,\theta)}{A(s,\theta)},
\end{align}
with $e(t)$ Gaussian white noise satisfying $\mathbb{E}[e(t)]=0$ and
$\mathbb{E}[e(t)e(s)]=\lambda\,\delta(t-s)$.  The monic polynomials are
\begin{align}
  A(s,\theta) &= s^n + a_1 s^{n-1} + \cdots + a_n,\\
  B(s,\theta) &= s^m + b_1 s^{m-1} + \cdots + b_m,\quad m < n,
\end{align}
Let $\eta:=\log\lambda$ and use the unconstrained scale coordinate
$\theta = [a_1,\dots,a_n,\,b_1,\dots,b_m,\,\eta]^T$, with
$\lambda=e^\eta$ the physical noise intensity reported in the numerical
results.
The corresponding power spectral density is
\begin{equation}
  \Phi(j\omega,\theta)
    = e^\eta\,\biggl|\frac{B(j\omega,\theta)}{A(j\omega,\theta)}\biggr|^{\!2}.
\end{equation}
The score gradient $s_k(\theta)=\nabla_\theta\log\Phi(j\omega_k,\theta)$
required by Algorithm~\robustalgorithmref\ takes the explicit form
\begin{equation}\label{eq:carmagrad}
\begin{aligned}
  \frac{\partial\log\Phi}{\partial a_\ell}
    &= -2\,\mathrm{Re}\!\left[\frac{(j\omega)^{n-\ell}}{A(j\omega,\theta)}\right],
    \quad \ell=1,\dots,n, \\
  \frac{\partial\log\Phi}{\partial b_\ell}
    &= \phantom{-}2\,\mathrm{Re}\!\left[\frac{(j\omega)^{m-\ell}}{B(j\omega,\theta)}\right],
    \quad \ell=1,\dots,m, \\
  \frac{\partial\log\Phi}{\partial\eta}
    &= 1.
\end{aligned}
\end{equation}
These are evaluated at $\omega=\omega_k$ for each DFT frequency and
used to evaluate the robust criterion gradient and the score-residual
check in Algorithm~\robustalgorithmref.

For an observation window of length $T$, the energy-normalised Fourier
transform and periodogram are
\begin{align}
  Y_T(j\omega)  &= \frac{1}{\sqrt{T}}\int_0^T y(t)\,e^{-j\omega t}\,dt,\\
  I_{N_s}(\omega_k)   &:= |Y_T(j\omega_k)|^2.
\end{align}
The $1/\sqrt{T}$ normalisation ensures
$\mathbb{E}[I_{N_s}(\omega_k)]\to\Phi(j\omega_k,\theta_0)$ as
$T\to\infty$, matching the continuous-time PSD convention
\cite{pintelon12}.
At the DFT frequencies $\omega_k = 2\pi k/T$, $k=1,\dots,N$
(excluding DC and, when $N_s$ is even, the Nyquist frequency
$\omega_{N_s/2}=\pi/\Delta t$, since both $Y_T(0)$ and
$Y_T(j\omega_{N_s/2})$ are real-valued and do not satisfy the
$\mathcal{CN}$ assumption), the
normalised residuals are asymptotically independent with
\begin{equation}
  \rho_k(\theta_0) = \frac{I_{N_s}(\omega_k)}{\Phi(j\omega_k,\theta_0)}
  \;\overset{a}{\sim}\; \mathrm{Exp}(1),
\end{equation}
since for large $T$ the DFT $Y_T(j\omega_k)$ is approximately
$\mathcal{CN}(0,\Phi(j\omega_k,\theta_0))$
\cite{brillinger81,dzhaparidze70,stoica05}; dividing the
squared modulus of a $\mathcal{CN}(0,\sigma^2)$ variable by
$\sigma^2$ yields an $\mathrm{Exp}(1)$ variate.
Although the periodogram itself is an inconsistent estimator of the
spectrum \cite{brillinger81}, the normalised ratio $\rho_k$ converges
in distribution, which is the property exploited below.

\subsection{Sampled-Data Implementation}
Suppose $y[n]=y(n\Delta t)$ is observed at a fixed sampling interval and
write $\nu_k=2\pi k/N_s$ in rad/sample.  The periodogram used for
estimation is
\begin{equation}\label{eq:sampledperiodogram}
  I_{N_s}(\nu_k)=\frac{1}{N_s}
    \Bigl|\sum_{n=0}^{N_s-1}y[n]e^{-j\nu_k n}\Bigr|^2,
  \qquad \omega_k=\nu_k/\Delta t.
\end{equation}
Its model is the exact sampled-process spectrum, not the unaliased CT
spectrum.  For ideal instantaneous sampling the scalar alias representation is
\begin{equation}\label{eq:sampled-spectrum}
  f_{\Delta}(\nu,\theta)=\frac{1}{\Delta t}
  \sum_{\ell\in\mathbb Z}
  \Phi\!\left(j\frac{\nu+2\pi\ell}{\Delta t},\theta\right),
  \qquad -\pi\leq\nu\leq\pi.
\end{equation}
Equivalently, for a CT state-space realisation $(A,C,L)$,
$F_{\Delta}=e^{A\Delta t}$ and
$Q_{\Delta}=\int_0^{\Delta t}e^{Au}LL^Te^{A^Tu}\,du$ give
$f_{\Delta}(\nu)=C(I-F_{\Delta}e^{-j\nu})^{-1}Q_{\Delta}
(I-F_{\Delta}^Te^{j\nu})^{-1}C^T$.
Any known analogue anti-alias filter is included inside the pre-sampling
spectrum in \eqref{eq:sampled-spectrum}.

At fixed $\Delta t$, the long-span regime is $N_s\to\infty$ and
$T=N_s\Delta t\to\infty$.  Under the Gaussian, stability, mixing,
smoothness and sampled-spectrum identifiability conditions stated in
Appendix~\ref{app:lemmas}, distinct positive-frequency DFT ordinates have
the usual asymptotic complex-normal law and
\[
  \rho_k(\theta_0):=I_{N_s}(\nu_k)/f_{\Delta}(\nu_k,\theta_0)
  \ \overset{a}{\sim}\ \operatorname{Exp}(1).
\]
This is the route by which the robust Whittle results apply to sampled CT
models \cite{fasenMayer22}.  A joint limit with $\Delta t\to0$ is a
different problem and is not claimed here \cite{brockwell12}.  The present
paper treats output-only identification; input-output extensions require a
separate input, noise and transient model and are outside its scope.

Below, $\Phi_k(\theta)$ denotes the spectrum actually used at ordinate~$k$:
$\Phi(j\omega_k,\theta)$ for ideal continuous observation and
$f_{\Delta}(\nu_k,\theta)$ for sampled data.  The definitions of
$\rho_k=I_k/\Phi_k$ and $s_k=\nabla_\theta\log\Phi_k$ use this convention.

For the discrete-time AR(2) proxy used in Section~\ref{sec:example},
the poles lie at $r e^{\pm j\phi}$ with $a_1=-2r\cos\phi$ and $a_2=r^2$,
so $r=\sqrt{a_2}$ is the pole radius and $\phi=\arccos(-a_1/(2r))$ the
resonant frequency.

\subsection*{Notation}

\small\begin{tabular}{@{}lp{0.48\columnwidth}@{}}
  $j = \sqrt{-1}$                   & imaginary unit \\
  $N_s$                              & number of time-domain samples \\
  $N = \lfloor N_s/2\rfloor$        & number of DFT frequencies \\
  $T = N_s\,\Delta t$               & observation time \\
  $\Phi_k(\theta)$                    & applicable CT or exact sampled model PSD \\
  $I_{N_s}(\nu_k)$                    & sampled periodogram at DFT frequency $\nu_k$ \\
  $\rho_k(\theta)$                    & $I_k/\Phi_k(\theta)$ \\
  $r_k(\theta)$                      & $\log\rho_k(\theta)$, log-spectral residual
                                       (cf.\ $\varepsilon(t,\theta)$ in \cite{ljung99}) \\
  $s_k(\theta)$                      & $\nabla_\theta\log\Phi_k(\theta)$, log-spectral gradient
                                       (carrier; cf.\ $x_i$ in \cite{maronna06}) \\
  $I(\theta)$                        & $\tfrac{1}{N}\!\sum_k s_k s_k^T$ \\
  $q_{\mathrm{W}}(x)=x-1$            & Whittle (maximum-likelihood, ML) score \\
  $q_b(\rho_k)=\psi_b(\log\rho_k)$  & robust score in the log-ratio coordinate \\
  $K(q)$                             & asymptotic variance scaling factor for score $q$ \\
  $\mathcal{CN}(0,\Sigma)$           & complex circular Gaussian \\
  $\mathrm{Exp}(1)$                  & exponential, density $e^{-x}$, $x\!\ge\! 0$
\end{tabular}\normalsize

\medskip

\section{Robust Frequency-Domain Estimation}\label{sec:robust}

The Whittle criterion is the natural frequency-domain ML estimator, but
a single outlying periodogram ordinate can dominate the entire loss.
The question is how to bound this influence without sacrificing the
statistical efficiency that makes ML attractive.  The answer comes from
a structural property of the Whittle residuals: a logarithmic
transformation reveals a Gumbel location model.  Classical local minimax
theory then yields an explicit optimal score, an integrable loss, and
a practical algorithm with provable consistency and efficiency
guarantees.

The Whittle criterion \cite{whittle61,dzhaparidze70,ljung99} can be formulated as
\begin{align}
  L_{N}(\theta) &
= \sum_{k=1}^{N}
    \biggl[\frac{I_k}{\Phi_k(\theta)}
    +\log \Phi_k(\theta)\biggr],  \\
  \hat\theta_{N} &= \arg\min_\theta L_{N}(\theta).
\end{align}
Because adding another constant does not change the solution, we can equivalently minimise
\begin{align}
  L_{N}(\theta) & =  \sum_{k=1}^{N} \rho_k(\theta) - \log\rho_k(\theta)
    =  \sum_{k=1}^{N} e^{r_k(\theta)} - r_k(\theta) \\
   & =  \sum_{k=1}^{N} \ell_{\mathrm{W}}(r_k(\theta)),
\end{align}
following the robust norm notation of \cite{ljung99}, with
$r_k(\theta) := \log\rho_k(\theta)$ the log-spectral residual,
the frequency-domain analogue of the prediction error
$\varepsilon(t,\theta)$ in \cite{ljung99}.
The Whittle criterion is a pseudo-likelihood based on the asymptotic
Gaussian distribution of the periodogram; at the nominal model it is
consistent and asymptotically efficient \cite{wald49,cramer46}.  When narrow-band disturbances are present,
however, outlying periodogram ordinates inflate $L_{N}(\theta)$
and introduce bias: because the loss sums over all frequencies, a
single large ordinate at a high-leverage frequency (near a spectral
resonance, for example) can dominate the gradient and pull the
entire parameter estimate toward it.

\begin{rmk}[Information-theoretic interpretation]
The per-frequency Whittle loss $\rho_k - \log\rho_k - 1 = e^{r_k} - r_k - 1$
is the Kullback--Leibler divergence between the empirical spectral
distribution $\mathcal{CN}(0, I_k)$ and the model
$\mathcal{CN}(0, \Phi_k(\theta))$ at frequency~$k$,
equivalently, the Itakura--Saito divergence between the periodogram
and model PSD\@.  The Whittle criterion thus minimises the total spectral
KL divergence.  The robust variant \eqref{eq:robust-loss-sum} bounds the
derivative of each per-frequency contribution, rather than the loss value
itself, preventing any single frequency from exerting unbounded influence
on the estimating equations.
\end{rmk}

\subsection{M-estimation}
The classical remedy for outlier sensitivity is \emph{M-estimation}
\cite{huber81,maronna06}: instead of minimising a sum of a fixed
loss~$\ell_{\mathrm{W}}$, one replaces $\ell_{\mathrm{W}}$ with a
loss~$\ell_b$ that grows more slowly in the tails, thereby limiting
the influence of any single observation on the estimate.

Equivalently, one solves a set of \emph{score equations}
\begin{equation}
\sum_k \psi(r_k(\theta))\,\nabla_\theta r_k(\theta) = 0,
\end{equation}
where the score~$\psi = \ell_b'$ downweights or clips extreme residuals.
The ``M'' stands for ``maximum-likelihood type'': ML is the special
case $\ell_b = \ell_{\mathrm{W}}$.  A bounded, centred score satisfying
the population-identification and regularity conditions below yields a
consistent, asymptotically normal estimator whose nominal variance overhead
relative to ML is quantified by a scalar factor~$K \geq 1$
(Theorem~\ref{thm:asympvar}).
The design question is which score to use.  The answer turns out to be
a clean one: the Whittle residuals possess a distributional structure
that reduces this problem to a classical one solved decades ago.

\subsection{Gumbel location formulation}
A logarithmic transformation converts the Whittle scale problem into a
Gumbel location problem, bringing classical local minimax robust estimation
theory directly to bear.  Setting
\begin{equation}\label{eq:gumbel-transform}
  r_k(\theta) = \log\rho_k(\theta),
\end{equation}
we find that under the nominal model $r_k$ has the standard (minimum)
Gumbel density $f_G(z) = \exp(z - e^z)$ on $\mathbb{R}$.  The Whittle
loss $\ell_{\mathrm{W}}(\rho) = \rho - \log\rho$ becomes the Gumbel
location loss $\ell_G(z) = e^z - z$ with location score
\begin{equation}\label{eq:gumbel-score}
  \ell_G'(z) = e^z - 1.
\end{equation}
The score is bounded below by $-1$ as $z\to-\infty$ but unbounded above,
which is what makes the Whittle estimator sensitive to outliers.
Back-transforming via $\rho = e^z$ recovers the Whittle ML score
$\rho - 1$ in the $\rho$-domain.  This equivalence between robust
Whittle estimation and robust Gumbel location estimation is the
structural insight we exploit in what follows.  It connects our
construction to the minimax theory of Rieder~\cite{rieder94} and to
the exponential-scale/Gumbel-location duality identified by
Thall~\cite{thall79} and treated computationally by Kohl~\cite{kohl05}.

\subsection{Robust Score via the Gumbel Model}

Clipping the Gumbel location score at level~$b$ and centering for
Fisher consistency yields the two-sided clipped Gumbel score,
equivalently the Rieder--Hampel bounded-influence score \cite{rieder94}.
This is the central object of the paper.  To see how it arises,
differentiating $L_{N}(\theta)$ yields the ML score equations
\begin{equation}\label{eq:mlscore}
  \sum_{k=1}^{N} \nabla_\theta r_k(\theta) \,\bigl(e^{r_k(\theta)} - 1\bigr) = 0.
\end{equation}
The Gumbel score $e^{r_k}-1$ grows exponentially for large~$r_k$,
so a sinusoidal disturbance ($r_k\gg 0$ at affected frequencies)
produces unbounded influence that biases~$\hat\theta_{N}$.

Robustness requires bounding this score on both sides.  Clipping the
Gumbel score $e^z - 1$ symmetrically at level~$b$ and adding a centering
correction $\alpha_b$ that ensures Fisher consistency
$E_{f_G}[\psi_b(z)] = 0$ yields
\begin{equation}\label{eq:gumbel-psi}
  \psi_b(z) = \mathrm{clip}\!\bigl(e^z - 1 + \alpha_b,\;-b,\;b\bigr).
\end{equation}
Transforming back via $\rho = e^z$ gives the $\rho$-domain score
\begin{equation}\label{eq:huberfunction}
  q_b(\rho) \;:=\; \psi_b(\log\rho) \;=\;
   \mathrm{clip}\bigl(\rho - m_b,\;-b,\;b\bigr),
\end{equation}
with $m_b := 1 - \alpha_b$ the centre of the affine middle band, and
$\alpha_b$ the unique centering correction satisfying
\begin{equation}\label{eq:centering}
  \mathbb{E}[q_b(R)] = 0, \qquad R \sim\mathrm{Exp}(1).
\end{equation}
Writing the threshold pair on the $\rho$-axis as
$\rho_{-} := \max(0,\,m_b - b)$ and $\rho_{+} := m_b + b$, the score
takes the value $-b$ on $[0,\rho_{-})$ (lower clip), the affine value
$\rho - m_b$ on $[\rho_{-},\rho_{+}]$, and $+b$ on $(\rho_{+},\infty)$
(upper clip).
Geometrically, clipping shifts the score's mean away from zero; the
correction $\alpha_b$ pulls it back just enough to restore unbiasedness,
ensuring Fisher consistency at~$\theta_0$.
The lower clip activates only when $b$ is small enough that $\rho_{-}>0$;
this happens for $b<b_0$ where $b_0=1+\tfrac12 W_0(-2 e^{-2})\approx 0.797$
(with $W_0$ the principal branch of Lambert's $W$ function) is the
unique positive solution of $1-b=e^{-2b}$.
For $b\geq b_0$ we have $\rho_{-}=0$, the lower clip is inactive on the
support of $\mathrm{Exp}(1)$, and \eqref{eq:huberfunction} reduces
pointwise to the one-sided form
$\min(\rho - 1 + \alpha_b,\;b)$ used throughout the practitioner range
$b\in[0.8,2.0]$.
The correction $\alpha_b$ ensures the estimating equations
\begin{align}\label{eq:estequation}
  \sum_{k=1}^{N}  \nabla_\theta \log \rho_k(\theta)\,q_b(\rho_k(\theta)) = 0
\end{align}
are unbiased at~$\theta_0$ for any $b > 0$
(Fig.~\ref{fig:psi}).

\subsection{The Robust Whittle Loss}

\begin{figure}[t]
  \centering
  \begin{subfigure}{0.48\columnwidth}
    \centering
    \includegraphics[width=\linewidth]{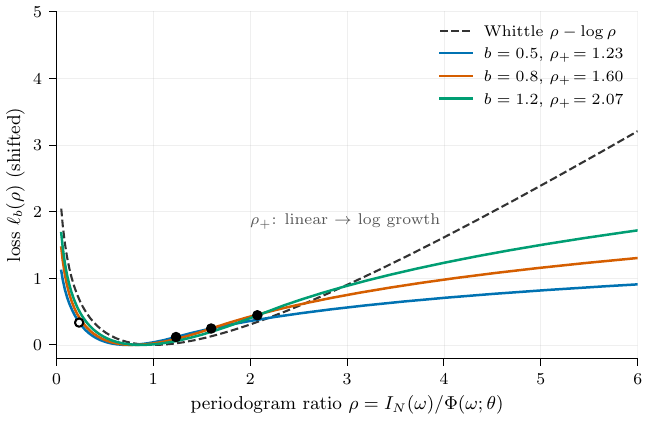}
    \caption{Loss $\ell_b(\rho)$ \eqref{eq:robust-loss}.}\label{fig:loss}
  \end{subfigure}
  \hfill
  \begin{subfigure}{0.48\columnwidth}
    \centering
    \includegraphics[width=\linewidth]{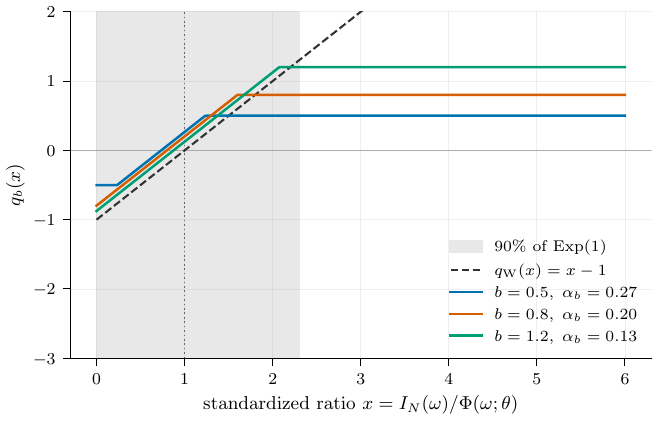}
    \caption{Log-ratio score $q_b(\rho)$ \eqref{eq:huberfunction}.}\label{fig:psi}
  \end{subfigure}
  \caption{Robust Whittle loss~(a) and corresponding score~(b),
    both shown for $b\in\{0.5,\,0.8,\,1.2\}$ (solid/coloured) versus
    the standard Whittle/ML reference $\rho-\log\rho$, $x-1$
    (dashed).  Dots in~(a) mark the upper transition
    $\rho_{+}=m_b+b$; beyond this point the loss grows as
    $b\log\rho$ (bounded influence).  Grey shading in~(b): central
    90\% of $\mathrm{Exp}(1)$.  For $b<b_0\approx 0.797$ the lower
    threshold $\rho_{-}=m_b-b$ is positive (e.g.\ $\rho_{-}\approx
    0.23$ at $b=0.5$); for $b\geq b_0$ the lower clip is inactive
    on the support of $\mathrm{Exp}(1)$ and both the loss and the
    score reduce to the one-sided clipped form.  Larger~$b$ reduces
    clipping; $b\to\infty$ recovers ML.}
\end{figure}

Interior stationary points satisfy the score equations
\eqref{eq:estequation}.  An explicit loss function, useful for monitoring
convergence and for comparison with the standard Whittle criterion,
is obtained by integrating the score \eqref{eq:gumbel-psi}.
Integrating $\psi_b(z)$ in the $z$-domain and transforming back via
$z = \log\rho$ gives, by the chain rule,
$d\ell_b/d\rho = \psi_b(\log\rho)/\rho = q_b(\rho)/\rho$.
Integrating in $\rho$ yields, on the three regions
$[0,\rho_{-})$, $[\rho_{-},\rho_{+}]$, $(\rho_{+},\infty)$:
\begin{equation}\label{eq:robust-loss}
  \ell_b(\rho) = \begin{cases}
    -b\log\dfrac{\rho}{\rho_{-}} + C_{-}
    & \rho < \rho_{-}, \\[4pt]
    \rho - (1{-}\alpha_b)\log\rho
    & \rho_{-}\leq \rho \leq \rho_{+}, \\[4pt]
    b\log\dfrac{\rho}{\rho_{+}} + C_{+}
    & \rho > \rho_{+},
  \end{cases}
\end{equation}
with the seam constants
$C_{-} := \rho_{-} - (1-\alpha_b)\log\rho_{-}$ and
$C_{+} := \rho_{+} - (1-\alpha_b)\log\rho_{+}$ chosen so the three
pieces agree at $\rho = \rho_{-}$ and $\rho = \rho_{+}$ (the additive
constant has no effect on the M-estimator).
For $b\geq b_0$ the lower-clip region is empty ($\rho_{-}=0$) and the
loss reduces to the familiar two-piece form: linear-plus-logarithmic
on $[0,\rho_{+}]$ and logarithmic for $\rho>\rho_{+}$.
On the affine middle band the loss coincides with the Whittle loss
up to a logarithmic correction $\alpha_b\log\rho$ that vanishes as
$b\to\infty$; outside the affine band the linear growth in $\rho$ is
replaced by logarithmic growth $b\log\rho$, bounding the influence
of both large and small periodogram ordinates.  In the $z$-domain the
loss is convex (strictly so on the affine middle, since
$d^2\ell_b(e^z)/dz^2=e^z>0$; linear in $z$ on either clipped flat).
Figure~\ref{fig:loss} illustrates the effect: for small~$b$ the
loss flattens early, strongly limiting the influence of large
$\rho$; as $b\to\infty$ the \emph{robust Whittle criterion}
\begin{equation}\label{eq:robust-loss-sum}
 L_{N}^{(b)}(\theta) \;:=\; \sum_{k=1}^{N}\ell_b(\rho_k(\theta))
\end{equation}
converges to the standard Whittle loss $L_{N}(\theta)$ up to a
data-dependent additive constant.  The theoretical estimator is an
approximate global minimiser,
\begin{equation}\label{eq:robust-argmin}
  L_N^{(b)}(\hat\theta_N)
  \leq \inf_{\theta\in\Theta}L_N^{(b)}(\theta)+o_p(N).
\end{equation}
Every differentiable interior minimiser satisfies the estimating
equations~\eqref{eq:estequation}; defining the estimator through
\eqref{eq:robust-argmin}, rather than as an arbitrary score root, is what
supports the global consistency argument below.

\iflatexml
\medskip
\noindent\textbf{Algorithm 1: Robust Whittle Estimator.}\par
\else
\begin{algorithm}
\caption{Robust Whittle Estimator}
\label{alg:robust-est}
\fi
\noindent\textbf{Input:} periodogram $\{I_{N_s}(\omega_k)\}_{k=1}^{N}$,
applicable spectral model $\Phi_k(\theta)$,
clipping level $b$, centering $\alpha_b$ and thresholds
$\rho_{-},\rho_{+}$ (Fig.~\ref{fig:acurve}).
\begin{enumerate}
  \item \textbf{Parameterise stability.}  Optimise in coordinates that
        remain inside the stable-model set.  For the AR(2) implementation,
        $a_2=\tanh u_2$ and
        $a_1=(1+a_2)\tanh u_1$ parameterise the Schur-stable triangle.
  \item \textbf{Initialise} from both a fixed neutral point and the ML
        Whittle estimate.
  \item \textbf{Minimise} $L_N^{(b)}$ in the stable coordinates by
        L-BFGS-B \cite{nocedal06}.  Retain only finite, stable, converged
        candidates whose score norm satisfies the prescribed tolerance.
  \item \textbf{Select} the verified candidate with the smallest value of
        the criterion~\eqref{eq:robust-loss-sum}.  If no verified candidate
        remains, report solver failure rather than returning an initial or
        ML estimate as a robust estimate.
\end{enumerate}
\iflatexml
\medskip
\else
\end{algorithm}
\fi

\noindent\textit{Convergence.}
The robust objective $L_{N}^{(b)}(\theta)$ in \eqref{eq:robust-loss-sum} is non-convex in~$\theta$
(as is the Whittle likelihood), so only local convergence is guaranteed.
Under Conditions~1--5 the population loss is locally strongly convex
near~$\theta_0$ because its Hessian is $c_b I(\theta_0)\succ0$.
Corresponding finite-sample local curvature holds with probability tending
to one under the stated uniform differentiation condition, but the realised
criterion can still have other local minima.
Multiple starts reduce, but do not eliminate, the risk of selecting a
non-global local minimum in a finite sample.  The consistency result below
applies to an approximate global minimiser in the sense of
\eqref{eq:robust-argmin}; numerical convergence and score residuals are
therefore recorded in the simulation artefacts rather than assumed.
Each objective evaluation costs $O(Np)$ after the initial
$O(N_s\log N_s)$ FFT.

\subsection{Influence Function and Consistency}

\begin{figure}[t]
  \centering
  \includegraphics[width=\columnwidth]{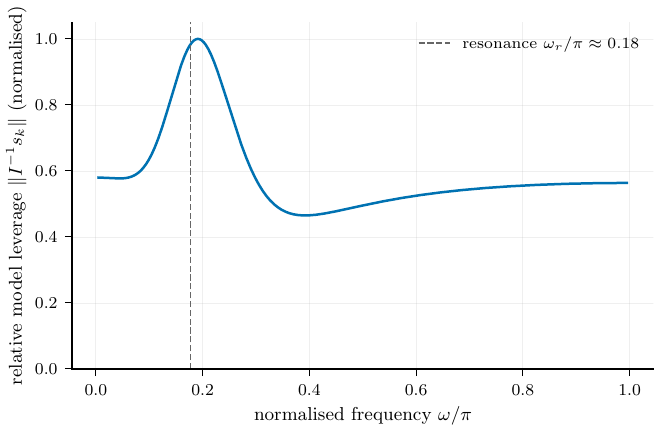}
  \caption{Model leverage $\|I(\theta)^{-1}s_k(\theta)\|$ for the
    AR(2) model, normalised to unit peak.  At fixed $b$, multiplication by
    $b/c_b$ gives the frequency-wise gross-error bound.
    The sensitivity peaks at the resonance
    $\omega_r/\pi\approx 0.18$ (dashed): high-leverage ordinates near
    the spectral peak are the primary targets for robust
    down-weighting.}
  \label{fig:relsensitivity}
\end{figure}

The bounded score ensures that no single periodogram ordinate can
have unbounded influence on the estimate.
The key quantities are
$s_k(\theta) = \nabla_\theta\log\Phi_k(\theta)$, the
log-spectral gradient, and the empirical information matrix
$I(\theta) = \frac{1}{N}\sum_k s_k s_k^T$ (the
frequency-domain analogue of $X^TX/n$ in regression).
Note that $s_k$ depends only on the parametric model, not on the
data.  Since $\nabla_\theta\log\rho_k = -s_k$, it controls the
direction in which each frequency can push the estimate.
Linearising the estimating equations \eqref{eq:estequation}
at the nominal model gives the bread matrix $-c_b I(\theta_0)$, where
\[
  c_b:=E[Rq_b'(R)]>0,\qquad R\sim\operatorname{Exp}(1).
\]
Thus, following
\cite{ljung99}~eq.~(15.12), the sensitivity of $\hat\theta$ to
ordinate~$k$:
\begin{equation}\label{eq:sensitivity}
  S_k(\rho,\theta) :=
  \frac{1}{c_b}I(\theta)^{-1}s_k(\theta)\,q_b(\rho).
\end{equation}
Here $I(\theta)^{-1}s_k(\theta)$ is the direction in
parameter space along which frequency~$\omega_k$ acts, determined
entirely by the model.  The normalised score $q_b(\rho)/c_b$ determines
how far the data at that frequency push.  Because $|q_b|\leq b$, the
influence is bounded:
\begin{equation}
  \sup_\rho \|S_k(\rho,\theta)\| \le
  \frac{b}{c_b}\,\|I(\theta)^{-1}s_k(\theta)\|
    =: \Gamma_k(\theta;b).
\end{equation}
The quantity $\Gamma_k=(b/c_b)\|I^{-1}s_k\|$ is the frequency-wise
gross-error bound \cite{hampel86}.  Its frequency dependence is carried by
the model-leverage factor $\|I^{-1}s_k\|$.  The hat-matrix diagonal
$h_{kk}=x_k^T(X^TX)^{-1}x_k$ identifies observations that can move a
least-squares fit regardless of their residual value; $\|I^{-1}s_k\|$
does the same for periodogram ordinates, identifying frequencies at
which a single outlier exerts the largest leverage on $\hat\theta$.
These frequencies are the primary targets for robust down-weighting
(Fig.~\ref{fig:relsensitivity}).

Before studying efficiency, we verify that an approximate global minimiser
of~\eqref{eq:robust-loss-sum} converges to the true parameter.  Under the
ideal nominal model, write
\[
  q_\theta(\xi):=
  \frac{f_{\theta_0}(\xi)}{f_\theta(\xi)},
  \qquad R\sim\operatorname{Exp}(1),
\]
where $f_\theta$ is the applicable CT or sampled spectrum, so that the
residual at frequency $\xi$ has the limiting form $q_\theta(\xi)R$.
Define the scalar population loss and score
\[
  M_b(q):=E[\ell_b(qR)],
  \qquad g_b(q):=E[q_b(qR)], \qquad q>0.
\]
Differentiation with respect to $\log q$ gives
\begin{equation}\label{eq:scalar-population-derivative}
  \frac{d}{d\log q}M_b(q)=g_b(q).
\end{equation}
The clipped score $q_b$ is non-decreasing and is strictly increasing on
its affine middle band.  Since $R$ has a positive density on $(0,\infty)$,
$g_b(q)$ is strictly increasing in $q$.  The centering condition gives
$g_b(1)=0$, so $M_b$ has its unique minimum at $q=1$.

Let $Q_b(\theta)$ be the limiting frequency average of
$E[\ell_b(q_\theta(\xi)R)]$.  The pointwise scalar result implies
$Q_b(\theta)\geq Q_b(\theta_0)$, with strict inequality whenever
$q_\theta(\xi)\ne1$ on a set of positive measure.  Condition~5 then
turns equality of spectra almost everywhere into $\theta=\theta_0$.
Thus spectral identifiability establishes a unique \emph{population
minimiser}; it is not being used to assert that the population score has no
other stationary zero.

Conditions~1, 3 and~4 give the uniform law of large numbers
\[
  \sup_{\theta\in\Theta}
  \left|N^{-1}L_N^{(b)}(\theta)-Q_b(\theta)\right|
  \xrightarrow{p}0.
\]
Compactness of $\Theta$, the unique-minimiser result, and the approximate
argmin condition~\eqref{eq:robust-argmin} now yield
$\hat\theta_N\xrightarrow{p}\theta_0$ by the M-estimator argmin theorem
\cite[Thm~5.7]{vandervaart98}.  Positive definiteness of
$I(\theta_0)$ (Condition~2) supplies the local non-degeneracy needed for
the subsequent asymptotic expansion.
With consistency established, the next question is the
\emph{rate} at which $\hat\theta$ converges and the shape of
its limiting distribution.  This is the subject of
Section~\ref{sec:asymptotic}.

\subsection{Asymptotic Normality and Efficiency}\label{sec:asymptotic}

Bounding the score protects against outliers; the next result
quantifies the cost to estimation accuracy.  Under the conditions below,
the robust estimator is $\sqrt{N}$-consistent and asymptotically normal, with a
covariance that differs from the ML Cramér--Rao bound by a single
scalar factor $K(q) \ge 1$.  This factor is the price of
robustness.  It depends only on the score $q$ and the nominal
$\mathrm{Exp}(1)$ distribution, not on the spectral model.

The Exp(1) and independence properties underlying
Theorem~\ref{thm:asympvar} are asymptotic in the observation
length~$T$ (Condition~3); the asymptotic distribution stated below is
exact within that limit.
\begin{thm}[Asymptotic normality]\label{thm:asympvar}
Let $q\in\mathcal{P}_b$ be an absolutely continuous score
(i.e.\ $|q|\le b$, $\mathbb{E}[q(\rho)]=0$
for $\rho\sim\mathrm{Exp}(1)$, and
$\mathbb{E}[\rho\,q'(\rho)]\neq 0$), and let $\ell_q$ satisfy
$d\ell_q/d\rho=q(\rho)/\rho$.  Let $\hat\theta$ be a consistent interior
approximate minimiser of $\sum_k\ell_q(\rho_k(\theta))$ that satisfies the
corresponding score equation with probability tending to one.  For
$q=q_b$, these are exactly \eqref{eq:robust-argmin} and
\eqref{eq:estequation}.  Under the regularity conditions of
Appendix~\ref{app:lemmas},
as $N\to\infty$ (at fixed $\Delta t$ in the sampled setting),
\begin{align}
  \sqrt{N}\,(\hat\theta - \theta_0)
    &\Rightarrow \mathcal{N}\bigl(0,\,K(q)\,I(\theta_0)^{-1}\bigr), \\
  K(q)
    &:= \frac{\mathbb{E}[q(\rho)^2]}
            {\bigl(\mathbb{E}[\rho\,q'(\rho)]\bigr)^2},
    \quad \rho\sim\mathrm{Exp}(1). \label{eq:kappa}
\end{align}
After influence-curve normalisation, the two-sided clipped Gumbel form
\eqref{eq:huberfunction} minimises nominal asymptotic variance at its
gross-error-sensitivity bound (Theorem~\ref{thm:opt-huber}).
\end{thm}

\begin{rmk}[Exact, asymptotic, and simulation-based statements]\label{rmk:evidence-status}
The centering correction $\alpha_b$, the clipping thresholds, and the
closed-form factor $K(b)$ are exact within the ideal model
$R\sim\operatorname{Exp}(1)$.  The periodogram law, asymptotic
independence of ordinates, $\sqrt{N}$ (equivalently $\sqrt{T}$ at fixed
sampling interval) normality, and covariance
$K(b)I(\theta_0)^{-1}$ are asymptotic statements.  Bias reduction,
finite-sample efficiency, the sample-size diagnostic, and the comparison
with periodogram trimming in Section~\ref{sec:example} are Monte Carlo findings
for the stated designs; they are not consequences of the asymptotic theorem
alone.
\end{rmk}

The scalar $K(q)\ge 1$ quantifies the efficiency cost of
robustness: for the ML score $q_{\mathrm{W}}(\rho)=\rho-1$ one has
$K=1$, so any clipping introduces $K>1$.
Indeed, integration by parts gives
$E[\rho q'(\rho)]=E[q(\rho)(\rho-1)]$; Cauchy--Schwarz and
$E[(\rho-1)^2]=1$ imply the inequality, with equality only for a score
proportional to $\rho-1$ almost surely.
$K(q)$ plays the same role as the variance scaling factor
$K(\ell, f_e)$ of \cite{ljung99}, eqs.~(15.1)--(15.2), but is
evaluated under the $\mathrm{Exp}(1)$ distribution of normalised
periodogram residuals rather than the Gaussian distribution of
time-domain prediction errors.

\begin{rmk}[Convergence rate]
The theorem is stated with $\sqrt{N}$ since $N$ approximately
independent DFT ordinates carry the information.  For fixed sampling
interval $\Delta t$, $N = \lfloor N_s/2\rfloor \approx T/(2\Delta t)$,
so $\sqrt{N}$-consistency is equivalent to $\sqrt{T}$-consistency up
to a constant, the frequency-domain analogue of the $\sqrt{N_s}$
rate in time-domain PEM \cite{ljung99}.
\end{rmk}

\begin{rmk}[Cramér-Rao interpretation]
The asymptotic covariance $K(q)\,I(\theta_0)^{-1}$ equals
$K(q)$ times the ML Cramér-Rao lower bound $I(\theta_0)^{-1}$,
so the robust estimator pays exactly a factor of $K(q)\ge 1$
in asymptotic efficiency relative to ML.  For the Huber choice at
$b=1.5$, $K\approx 1.16$: a 16\% nominal variance overhead in exchange for
bounded influence.
\end{rmk}

\begin{rmk}
The denominator of $K$ is $\mathbb{E}[\rho\,q'(\rho)]^2$, not
$\mathbb{E}[q'(\rho)]^2$ as in the classical Huber sandwich.  This
difference arises because the gradient of the estimating
equation \eqref{eq:estequation}
involves $\rho_k$: differentiating $s_kq(\rho_k)$ with respect to $\theta$
and using $\nabla_\theta \rho_k = -\rho_k\, s_k$ gives a factor of
$\rho_kq'(\rho_k)$, whose expectation under $\mathrm{Exp}(1)$ is
$\mathbb{E}[\rho\,q'(\rho)]$.  Since the ``meat''
$\mathbb{E}[q(\rho)^2]\,I(\theta_0)$ and ``bread''
$\mathbb{E}[\rho\,q'(\rho)]\,I(\theta_0)$ of the sandwich share the same
matrix $I(\theta_0)$, the covariance simplifies exactly to
$K(q)\,I(\theta_0)^{-1}$, preserving the Fisher information
matrix structure.
\end{rmk}

For the two-sided clipped Gumbel score \eqref{eq:huberfunction}, the
general expression \eqref{eq:kappa} reduces to an explicit closed form.
Writing $\rho_{-} := \max(0, m_b - b)$ and $\rho_{+} := m_b + b$ for the
two clipping thresholds:
\begin{equation}\label{eq:Kexplicit}
  K(b) = \frac{N_b}{D_b^{\,2}},
\end{equation}
with
\begin{align*}
  N_b &= b^2\bigl(1 - e^{-\rho_{-}}\bigr)\mathbf 1_{\rho_{-}>0}
        + \!\!\int_{\rho_{-}}^{\rho_{+}}\!(\rho - m_b)^2 e^{-\rho}\,d\rho \\
      &\quad + b^2 e^{-\rho_{+}}, \\
  D_b &= (\rho_{-}\!+\!1)\,e^{-\rho_{-}}
       - (\rho_{+}\!+\!1)\,e^{-\rho_{+}}.
\end{align*}
The middle integral evaluates in elementary terms via
$\int(\rho-m_b)^2 e^{-\rho}d\rho =
-e^{-\rho}\bigl[(\rho-m_b)^2+2(\rho-m_b)+2\bigr]$
(Appendix~\ref{app:kappa}), so $K(b)$ is a finite combination of
exponentials and polynomials in~$b$, with no numerical optimisation
required.  For $b\geq b_0$ the lower-clip term drops out
($\rho_{-}=0$) and \eqref{eq:Kexplicit} recovers the one-sided closed
form used pre-correction.  Table~\ref{tab:kappa} lists values for
representative clipping levels.

\begin{table}[t]
  \caption{Theoretical efficiency factor $K(b)$ and centering correction
    $\alpha_b$ for selected clipping levels.  $K=1$ is the ML bound; the practical
    window $b\in[1,2]$ incurs at most 30\% nominal asymptotic variance overhead.
    For $b\geq b_0\approx 0.797$ the values agree with the
    pre-correction one-sided table within published precision; only
    the row at $b=0.5<b_0$ shifts (from $K=1.639$ to $K=1.579$),
    reflecting the lower-clip contribution.}
  \label{tab:kappa}
  \vspace{6pt}
  \centering\small
  \begin{tabular}{@{}cccc@{}}
    \toprule
    $b$ & $\alpha_b$ & $K(b)$ & Overhead (\%) \\
    \midrule
    0.5           & $0.266$ & 1.579 & 57.9 \\
    0.8           & $0.202$ & 1.389 & 38.9 \\
    1.0           & $0.159$ & 1.295 & 29.5 \\
    1.2           & $0.126$ & 1.229 & 22.9 \\
    1.5           & $0.090$ & 1.162 & 16.2 \\
    2.0           & $0.052$ & 1.094 &  9.4 \\
    3.0           & $0.019$ & 1.034 &  3.4 \\
    $\infty$ (ML) & $0$     & 1.000 &  0   \\
    \bottomrule
  \end{tabular}
\end{table}

\subsection{Finite-Sample Behaviour and Robustness to Model Deviations}

The Exp(1) and independence properties of the normalised periodogram
ordinates $\rho_k(\theta_0)$ are asymptotic in the observation
length~$T$, as formalised in Condition~3 (Appendix~\ref{app:lemmas})
and consistent with the standard Whittle framework
\cite{brillinger81,dzhaparidze70,stoica05}.  In finite samples these
properties hold approximately; the two principal sources of deviation
from the ideal asymptotic regime are spectral leakage arising from
the rectangular DFT window and model mismatch between
$\Phi(j\omega,\theta)$ and the true spectrum.

The robust score clips every periodogram bin outside the affine
middle band $[\rho_{-},\rho_{+}]$, so moderate leakage from the
rectangular window (which spreads spike energy into a band of
adjacent bins) is handled automatically by the bounded-influence
mechanism:
each contaminated adjacent bin is itself clipped and therefore cannot
accumulate unbounded influence.  For severe leakage, data tapering
can be applied at the cost of frequency resolution, and the asymptotic
theory of Section~\ref{sec:asymptotic} carries over with a
leakage-adjusted information matrix (see also the leakage discussion
in Section~\ref{sec:discussion}).

The defining property of the estimator is that the score
$|q_b(\rho)| \le b$ is bounded on both sides regardless of the
exact distribution of~$\rho_k$.  Even when finite-sample effects
perturb the $\mathrm{Exp}(1)$ approximation (whether through spectral
leakage, model mismatch, or a finite observation window), no single
periodogram ordinate can dominate the estimating equations
\eqref{eq:estequation}.  This per-ordinate bound does not by itself control
accumulated bias across many affected frequencies or establish robustness
under model misspecification; those require separate analyses.

\section{Bounded-Influence Optimality}\label{sec:optimality}

Theorem~\ref{thm:asympvar} gives the asymptotic covariance for a consistent
interior estimator using a bounded, centred score under the stated
regularity conditions; the nominal variance factor~$K$ varies across
scores.  The remaining question is
which score extracts the most information from uncontaminated
observations while still bounding the influence of outliers.  The
answer is the two-sided clipped Gumbel form already derived above,
and it comes from classical minimax theory applied in the Gumbel
domain.  Let $Z\sim G$ have the standard Gumbel law and location score
$\Lambda(z):=e^z-1$.  Put
$\mathcal C:=\{\eta\in L_2(G):E_G[\eta]=0,\ E_G[\eta\Lambda]=1\}$ and
$\mathcal C_B:=\{\eta\in\mathcal C:\|\eta\|_\infty\leq B\}$; these
constraints fix centering and scale.  For $r\geq0$, set
$q_N:=r/\sqrt N$ and define
\[
  \mathcal U_N(G,r):=\{(1-q_N)G+q_NH:H\text{ any distribution}\}.
\]
This is the classical infinitesimal-contamination regime: the
neighbourhood contracts at the same inverse-square-root asymptotic scale on
which an asymptotically linear estimator has a non-degenerate limit.
Its maximal asymptotic mean squared error for
influence curve $\eta$ is
$\mathcal R_r(\eta):=E_G[\eta^2]+r^2\|\eta\|_\infty^2$, the sum of
nominal variance and maximal squared first-order contamination bias
\cite[Ch.~5]{rieder94}; see also
\cite[Secs.~1.2--1.3 and 5.2.1]{kohl05}.

For the centred clipped score $\psi_b$ in \eqref{eq:gumbel-psi}, write
$c_b:=E_G[\psi_b\Lambda]=E_G[\psi_b']>0$,
$\eta_b:=\psi_b/c_b$, $B_b:=b/c_b$, and
\begin{equation}\label{eq:local-radius}
  r_b^2:=b^{-1}E_G\!\left[\bigl(|\Lambda+\alpha_b|-b\bigr)_+\right].
\end{equation}
The identity for $c_b$ follows by integration by parts and equals
$E[\rho\,q_b'(\rho)]$ in \eqref{eq:kappa}.
Equation~\eqref{eq:local-radius} is the scalar clipping--radius relation
from this theory after normalising the clipped score by its central slope.

\begin{thm}[Local minimax optimality]\label{thm:opt-huber}
For every $b>0$, the normalised influence curve $\eta_b$ uniquely
minimises $E_G[\eta^2]$ over $\mathcal C_{B_b}$ and
$\mathcal R_{r_b}(\eta)$ over $\mathcal C$, up to $G$-null sets.
Consequently, the two-sided clipped Gumbel score gives the exact
solution of the local asymptotic influence-curve minimax problem
associated with $\mathcal U_N(G,r_b)$, with
$E_G[\eta_b^2]=E_G[\psi_b^2]/c_b^2=K(b)$.
\end{thm}

Scaling the estimating equation by $1/c_b$ changes neither its root nor
the estimator.  The result is exact within this local asymptotic
influence-curve problem; it makes no minimax claim for a fixed positive
contamination fraction and arbitrary $H$.  The fixed-fraction
experiments in Section~\ref{sec:example} are finite-sample robustness
stress tests.  Appendix~\ref{app:proof-opt-huber} gives a direct proof.

\section{Tuning and Practical Guidance}\label{sec:tuning}

The tuning parameter $b$ controls the bias-variance trade-off:
large $b$ approaches ML (efficient, not robust); small $b$ clips
more aggressively (robust, higher variance as quantified by $K(b)$
in Table~\ref{tab:kappa}).  The centering correction $\alpha_b$ is
piecewise:
\begin{equation}\label{eq:abrelation}
  \alpha_b =
  \begin{cases}
    \displaystyle 1 - b - \log\!\frac{1 - e^{-2b}}{b}
      & 0 < b < b_0, \\[6pt]
    \alpha_b = e^{-(1+b-\alpha_b)}
      & b \geq b_0.
  \end{cases}
\end{equation}
The implicit second-line equation has a unique solution
$\alpha_b\in(0,1)$ for every $b > 0$: writing
$g(\alpha)=\alpha-e^{-(1+b-\alpha)}$, the map is strictly increasing
on $[0,1]$ with $g(0)<0<g(1)$, so the zero $\alpha_b$ exists and is
unique; we take that solution on $b\geq b_0$.
The threshold $b_0 = 1+\tfrac12 W_0(-2 e^{-2}) \approx 0.797$ is the
unique positive solution of $1-b = e^{-2b}$ and is the value of $b$
at which the lower clip activates.  Both branches agree at $b=b_0$
with $\alpha_{b_0} = 1 - b_0 \approx 0.203$, so $\alpha_b$ is
continuous in $b$.
For $b\geq b_0$ the second-line implicit equation is the same
$\alpha_b = e^{-(1+b-\alpha_b)}$ that arises from the one-sided
construction, so the practitioner range $b\in[0.8,2.0]$ inherits the
familiar closed-form characterisation.  The closed form on $0<b<b_0$
extends the estimator to small $b$ via the standard Rieder--Hampel
two-sided construction \cite{rieder94}.
The centering is exponentially small for large $b$:
$\alpha_b < 0.06$ for $b\geq 2$ and $\alpha_b\to 0$ as $b\to\infty$.

\noindent\textit{Practical guidance.}
A reasonable default is $b=1.5$, giving $K\approx 1.16$ (16\%
variance overhead).  Cross-validation over a grid of $b$ with
held-out frequency blocks provides a possible data-driven diagnostic, but
its selection effect must be included in any subsequent inference.  There is no universal map
from a time-domain sinusoid amplitude~$A$ to~$b$: the
normalised ordinate also depends on $N_s$, leakage, local spectral level and
frequency alignment.  The amplitude sweeps below therefore keep $b$ fixed
and are reported as finite-design sensitivity studies rather than a tuning
theorem.
Plots of efficiency and clipped fraction as functions of~$b$ are
provided in the companion repository; the shaded band
$b\in[0.8,\,2.0]$ marks the recommended range.

\medskip\hrule\medskip
\noindent\textbf{Practitioner recipe.}
Given a periodogram $\{I_{N_s}(\omega_k)\}$ and a CARMA model
$\Phi_k(\theta)$ (the exact $f_\Delta$ for sampled data):
\begin{enumerate}[nosep,leftmargin=*]
  \item Compute the ML estimate $\hat\theta_{\mathrm{ML}}$ by minimising
        the Whittle criterion.
  \item Choose $b$ from the nominal efficiency budget; $b=1.5$ gives
        $K\!\approx\!1.16$ (16\% variance overhead).  If $b$ is selected
        from the data, account for that selection in uncertainty statements.
  \item Minimise the robust criterion from both the neutral and ML starts
        and retain a verified candidate (Algorithm~\robustalgorithmref).
  \item Read $K(b)$ from Table~\ref{tab:kappa} to obtain the
        asymptotic covariance $K(b)\,I(\hat\theta)^{-1}/N$.
\end{enumerate}
\medskip\hrule\medskip
\noindent A flowchart summarising this workflow is available in the
companion repository.

\section{Numerical Example}\label{sec:example}

To test the theory against ground truth, we use a resonant
AR(2).  This is a canonical resonant system where narrow-band
disturbances near the spectral peak exert maximum leverage on the
parameter estimates.  The model is
\begin{equation}
  y[n] + a_1 y[n-1] + a_2 y[n-2] = e[n], \quad e\sim\mathcal{N}(0,\lambda),
\end{equation}
with true parameters $a_1 = -1.2$, $a_2 = 0.5$, $\lambda = 1.0$.
The normalised periodogram ratio $\rho_k$ follows the
$\mathrm{Exp}(1)$ asymptotic distribution for both DT and CT spectral
models.  The DT AR(2) therefore exercises exactly the same statistical
mechanism (Gumbel location M-estimation of periodogram residuals), but it
is not itself a sampled-CT identification experiment.  A separate exact
sampled-CAR(2) check is reported below.
Setting $\Delta t=1\,\mathrm{s}$ maps the DT frequencies
$\omega\in[0,\pi]$ to physical frequencies in rad/s.
The poles lie at $z=r\,e^{\pm j\phi}$ with $r=\sqrt{a_2}\approx0.71$
and $\phi=\arccos(-a_1/(2r))\approx0.18\pi\,\mathrm{rad}$.
The exact map $s=\log(z)/\Delta t$ gives CT poles
$s\approx -0.34\pm j0.57$, corresponding to a CAR(2) with undamped
natural frequency $\omega_n\approx0.66\,\mathrm{rad/s}$ and damping
ratio $\zeta\approx0.53$ (a moderately damped resonance).
The power spectral density is
$\Phi(j\omega,\theta) = \lambda/|1+a_1 e^{-j\omega}+a_2 e^{-2j\omega}|^2$,
with its discrete log-spectrum gradient used in the robust score equations
\eqref{eq:estequation}.

\subsection{Simulation Setup}

For the bias--variance experiment we perform $N_{\mathrm{MC}}=200$
Monte Carlo runs of length $N_s=4096$.
Each run generates clean and disturbed realisations (three sinusoidal
disturbances at random frequencies in $(0.05,\,0.45)\times\pi$,
amplitude $A=2$).  The periodogram is computed without windowing to
preserve the $\mathrm{Exp}(1)$ basis.
All axes show normalised frequency $\omega/\pi\in[0,1]$
($\Delta t=1\,\mathrm{s}$, so $\omega/\pi=0.18$ corresponds to
$\omega_r\approx0.57\,\mathrm{rad/s}$).
The ML estimate uses L-BFGS-B from a fixed neutral start
$\theta^{(0)}=[-0.5,\,0.25,\,\eta=0]^T$; the robust estimate follows
Algorithm~\robustalgorithmref\ by minimising the robust criterion,
using the neutral and ML starts.  We scan $b\in[0.6,\,3.0]$ at 25 points.

\subsection{Results}

\begin{figure}[t]
  \centering
  \includegraphics[width=\columnwidth]{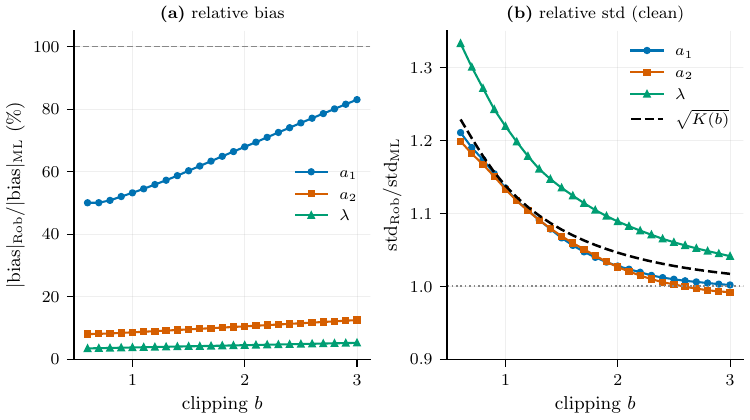}
  \caption{Bias--variance trade-off versus clipping level~$b$
    ($N_{\mathrm{MC}}=200$).
    (a)~Relative bias $|\mathrm{bias}|_{\mathrm{Rob}}/
    |\mathrm{bias}|_{\mathrm{ML}}$ on disturbed data ($A=2$): the
    reduction is parameter dependent, ranging over the displayed grid
    from $17$--$50\%$ for $a_1$, $87$--$92\%$ for $a_2$, and
    $95$--$97\%$ for $\lambda$.
    (b)~Relative standard deviation on clean data: the empirical ratios
    are compared with the asymptotic prediction $\sqrt{K(b)}$ (dashed).}
  \label{fig:relbias_robuststd}
\end{figure}

\paragraph*{Bias under disturbances.}
Figure~\ref{fig:relbias_robuststd}(a) shows relative bias versus $b$
on disturbed data ($A=2$).  The reduction is parameter dependent:
over $b\in[0.6,\,3.0]$ it ranges from $17$--$50\%$ for $a_1$,
$87$--$92\%$ for $a_2$, and $95$--$97\%$ for $\lambda$.

\paragraph*{Variance without disturbances.}
Figure~\ref{fig:relbias_robuststd}(b) shows the ratio of robust to ML
standard deviations on clean data.  The empirical curves broadly track
the theoretical $\sqrt{K(b)}$ (dashed); with 200 realisations, the median
absolute discrepancy over parameters and clipping levels is $0.020$.

\begin{figure}[t]
  \centering
  \includegraphics[width=\columnwidth]{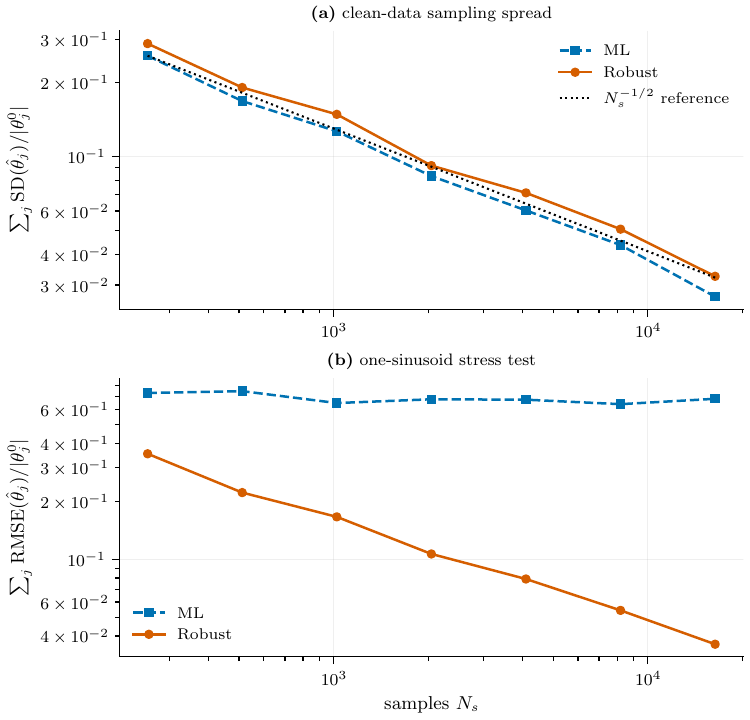}
  \caption{Sample-size diagnostic at $b=0.8$
    ($N_{\mathrm{MC}}=80$ per $N_s$).
    (a)~Sum of per-parameter relative standard deviations on clean data;
    the dotted line is an $N_s^{-1/2}$ reference anchored at the first
    ML point.  (b)~Sum of per-parameter relative RMSE under a single
    sinusoidal disturbance ($A=1.5$).  The robust RMSE decreases across
    the displayed range, while the ML RMSE remains approximately flat.
    These curves are Monte Carlo diagnostics for this design.}
  \label{fig:Ndep}
\end{figure}

\paragraph*{Sample-size diagnostic.}
On clean data, Fig.~\ref{fig:Ndep}(a) gives fitted log--log slopes
$-0.53$ for ML and $-0.51$ for the robust estimator, compatible with
the $N_s^{-1/2}$ asymptotic rate.  In the one-sinusoid finite-design stress
test, panel~(b) reports RMSE rather than standard deviation: the robust
RMSE has fitted slope $-0.54$, while the ML RMSE remains approximately
flat (slope $-0.02$).  The generator holds the number of sinusoids at one
and redraws its frequency in the documented range for each replicate; this
is not asserted to realise either a fixed-fraction or shrinking gross-error
neighbourhood.  The trends are diagnostics, not proofs of consistency.

\begin{figure}[t]
  \centering
  \includegraphics[width=\columnwidth]{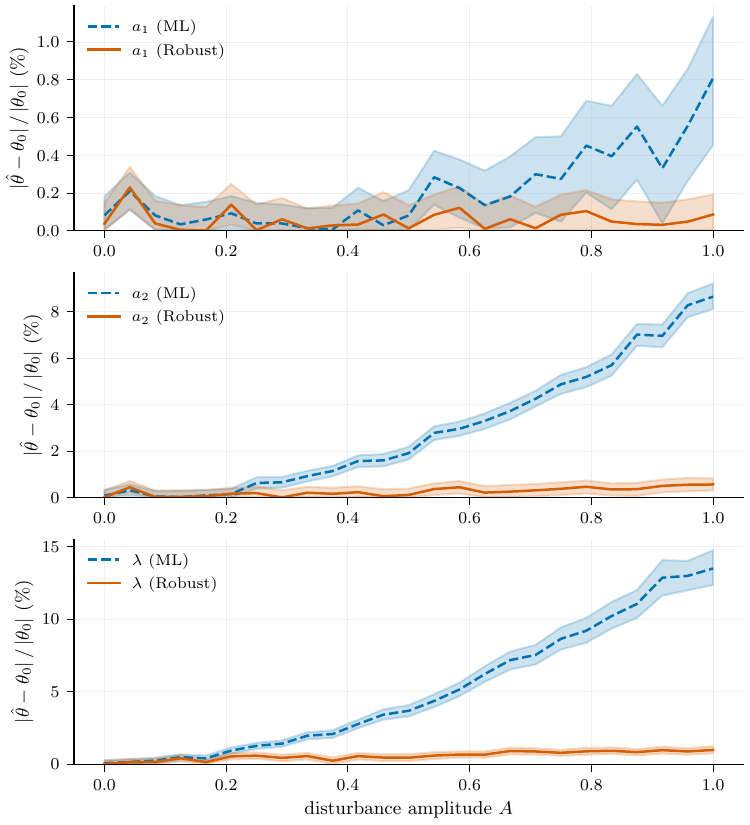}
  \caption{Per-parameter mean relative bias
    $|\hat{\theta}-\theta_0|/|\theta_0|$ (\%) versus disturbance
    amplitude~$A$ at $b=1.0$
    ($N_s=4096$, $N_{\mathrm{MC}}=500$).
    Dashed: ML; solid: robust.  Shaded bands show
    bootstrap 95\% confidence intervals on the mean.
    The ML bias grows steadily with~$A$, while the
    robust estimator remains near zero across the
    full amplitude range.}
  \label{fig:robustbias}
\end{figure}

\paragraph*{Bias versus disturbance amplitude.}
Figure~\ref{fig:robustbias} shows mean relative bias versus $A$ at
$b=1.0$: the ML bias grows steadily while the robust estimator
remains near zero.  Bootstrap 95\% bands confirm statistical precision.

\begin{figure}[t]
  \centering
  \includegraphics[width=\columnwidth]{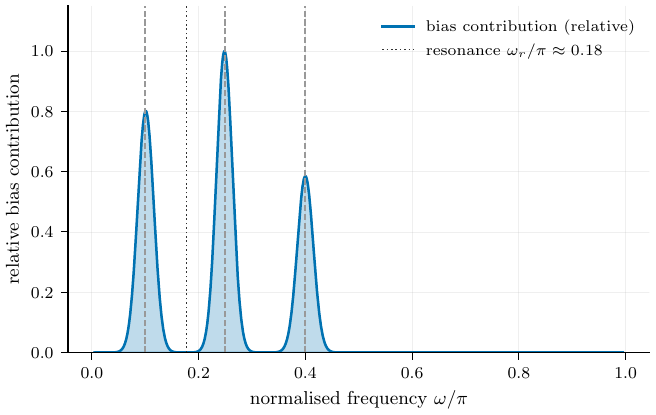}
  \caption{Frequency-wise relative bias contribution
    ($\|I^{-1}s_k\|\times$ disturbance profile, normalised) for
    three narrow-band spikes at $\omega/\pi\in\{0.10,\,0.25,\,0.40\}$
    (dashed verticals) and the resonance $\omega_r/\pi\approx 0.18$
    (dotted).  The spike nearest the spectral peak produces the
    largest contribution, motivating frequency-selective clipping
    with priority near $\omega_r$.}
  \label{fig:biascontrib}
\end{figure}

\paragraph*{Frequency-wise bias contribution.}
Figure~\ref{fig:biascontrib} shows $\|I^{-1}s_k\|\times$disturbance
profile for three spikes.  The contribution is largest near the
spectral resonance ($\omega_r/\pi\approx 0.18$), confirming that
frequency-selective clipping near the spectral peak yields the
greatest bias reduction.

\paragraph*{Exact sampled-CAR(2) check.}
To exercise the continuous-time identification path directly, a separate
Gaussian CAR(2) experiment uses
$(a_1,a_2,\lambda)=(0.7,2.0,1.3)$, $\Delta t=0.5$\,s and $N_s=4096$.
Each trajectory is generated from the exact discrete state transition
$F_\Delta=e^{A\Delta t}$ and covariance
$Q_\Delta=P-F_\Delta P F_\Delta^T$, and both estimators use the exact
sampled spectrum in \eqref{eq:sampled-spectrum}.  Forty seeded replicates
are run clean and with one sinusoid of amplitude~1.2 at
$\nu=0.75$\,rad/sample (seeded random phase); the robust fit uses $b=1.2$.
Table~\ref{tab:sampled-car2} reports relative RMSE.  All 160 optimisation
fits converged, with maximum reported absolute objective-gradient component
$2.3\times10^{-7}$.  These are finite-design Monte Carlo results, not a
proof of the asymptotic theorem or of universal superiority.

\begin{table}[t]
  \caption{Relative RMSE in the exact sampled-CAR(2) check
    ($N_{\mathrm{MC}}=40$).  Replicate-level estimates, seeds, source hashes
    and the complete protocol are frozen in the validation bundle.}
  \label{tab:sampled-car2}
  \vspace{4pt}
  \centering\small
  \begin{tabular}{@{}llccc@{}}
    \toprule
    Regime & Estimator & $a_1$ & $a_2$ & $\lambda$ \\
    \midrule
    Clean    & ML     & 0.041 & 0.018 & 0.027 \\
    Clean    & Robust & 0.043 & 0.022 & 0.027 \\
    One tone & ML     & 0.669 & 0.076 & 0.090 \\
    One tone & Robust & 0.049 & 0.023 & 0.027 \\
    \bottomrule
  \end{tabular}
\end{table}

\FloatBarrier

\subsection{Discussion}\label{sec:discussion}

The simulations illustrate the predicted bias--variance trade-off:
robust clipping reduces bias under narrow-band disturbances at a
moderate, analytically predictable asymptotic efficiency cost.  The
practical recommendation is
$b \in [1.0,\,1.5]$ (Table~\ref{tab:kappa}: $K \approx 1.16$--$1.30$,
i.e.\ 16--30\% variance overhead), which yields substantial bias reduction
while keeping the variance ratio near unity.  For stronger disturbances or
more stringent bias requirements, $b$ can be reduced further at the cost
of larger $K$.

\paragraph*{Comparison with periodogram trimming.}
The standard alternative is to remove periodogram bins around detected
spike frequencies before fitting.  This discards gradient information,
requires a location and bandwidth rule, and makes peak-picking errors part
of the estimator.  Oracle trimming at fixed locations leaves the nominal
Exp(1) model intact on the retained ordinates, but with a reduced information
sum; data-dependent trimming requires additional selection-aware analysis.
The robust estimator instead retains all ordinates with bounded influence.

\begin{figure}[t]
  \centering
  \includegraphics[width=\columnwidth]{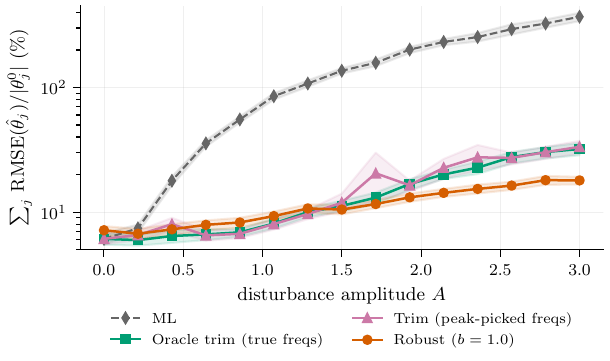}
  \caption{Sum of per-parameter relative RMSE,
    $\sum_j\mathrm{RMSE}(\hat\theta_j)/|\theta_j^0|$,
    versus disturbance amplitude~$A$ (logarithmic vertical scale)
    ($N_{\mathrm{MC}}=100$, $b=1.0$, three narrow-band spikes).
    RMSE captures both bias and variance, penalising methods
    that reduce bias at the cost of increased variability.
    Grey: unprotected ML.  Green: oracle trimming at true frequencies;
    purple: equal-budget trimming at peak-picked frequencies; orange:
    robust estimation.  For $A>0$ each trimming method excludes fifteen
    raw periodogram bins; $A=0$ is a shared no-trimming ML reference.
    Oracle trimming is below robust estimation for $A=0.21$--$1.29$ in
    the point estimates (peak-picked trimming is marginally lowest at some
    intermediate amplitudes), while robust estimation leads both trimming
    methods for $A\geq1.5$.
    Shaded bands show bootstrap 95\,\% CI\@.}
  \label{fig:notch}
\end{figure}

Figure~\ref{fig:notch} compares equal-budget trimming with ML and the
robust estimator.  For $A>0$, both trimming methods discard exactly fifteen
raw periodogram bins (five per disturbance in aggregate); no time-domain
filter is applied, and $A=0$ is a shared no-trimming ML reference.  In the
Monte Carlo point estimates, oracle trimming has lower aggregate RMSE than
the robust estimator for $A=0.21$--$1.29$.  The robust estimator has lower
RMSE than both trimming methods at every plotted amplitude $A\geq1.5$,
without requiring frequency identification.  Thus trimming is effective
when tones are weak and correctly located, whereas bounded influence is
less sensitive to leakage beyond a fixed exclusion band as amplitude grows.

\paragraph*{Alternatives.}
Trimmed periodograms can be effective when disturbance locations are
known, as Fig.~\ref{fig:notch} demonstrates, but require a detection and
bandwidth rule; data-dependent trimming also changes the inferential
problem.  Time-domain robust AR methods address a different contamination
geometry.  Neither alternative supplies the specific local Gumbel
influence-curve minimax characterisation and closed-form~$K(b)$ above.
Model-order selection and robustness under structural misspecification are
separate inferential problems and are not claimed as results of the present
finite-design study.

\subsection{Comparison with Related Approaches}\label{sec:related-comparison}

\paragraph*{Time-domain robust PEM.}
Classical time-domain robust identification applies Huber-loss
M-estimators to the prediction-error sequence
\cite{ljung99,maronna06}.  This approach downweights large
\emph{temporal} outliers.  A persistent narrow-band tone instead produces
structured residual dependence and concentrates energy spectrally.  This
motivates applying the bounded score to periodogram ratios.  The present
experiments do not constitute a matched comparison with robust PEM, whose
performance depends on its disturbance and predictor model.

\paragraph*{Heavy-tailed likelihood.}
Replacing the Gaussian innovation model with a heavier-tailed
distribution (e.g.\ Student-$t$ innovations) targets impulsive random
innovations.  A deterministic narrow-band tone is a different, structured
departure from that nominal model.  The proposed estimator uses only a
bounded periodogram-ratio score for this departure; no universal ranking
against heavy-tailed likelihoods is claimed \cite{kanakeri25}.

\paragraph*{Kernel-based identification.}
Kernel and regularised identification methods
\cite{pillonetto14,bottegal16}
regularise the impulse response via a Gaussian-process prior and have
been extended to robust variants that downweight outliers in the
time-domain residual sequence.  Their primary modelling target differs from
the parametric CARMA spectrum studied here, whose poles carry physical
interpretations.  Kernel methods may also be adapted to structured
disturbances; they are not compared numerically in this paper.

The approaches therefore emphasise different failure modes: temporal
outliers, heavy innovation tails, impulse-response regularisation and
spectrally concentrated disturbances.  The claimed contribution is the
last mechanism and its local Gumbel influence-curve analysis.

\paragraph*{Spectral leakage.}
The robust score bounds each affected bin, including bins reached by
leakage from the rectangular DFT window.  This limits per-bin influence but
does not remove leakage-induced dependence or accumulated bias.  Tapering
can be applied at the cost of resolution and requires its own covariance
accounting.

\paragraph*{Post-estimation diagnostics.}
A Q-Q plot of $\rho_k(\hat\theta)$ against Exp(1) provides a visual
check: outlying frequencies appear in the upper tail, while departures
at non-outlying frequencies signal model inadequacy.

\section{Conclusions}

Narrow-band disturbances occur in mechanical and electrical systems, and
the standard Whittle score gives an affected ordinate unbounded influence.
The framework developed here provides a bounded alternative: a single
logarithmic transformation reveals a Gumbel location structure that
connects robust spectral estimation to the classical minimax theory of
Huber and Rieder, yielding a two-sided clipped Gumbel score with a
piecewise closed-form centering correction whose normalised influence
curve is locally asymptotically minimax under shrinking gross-error
contamination.  Under the stated fixed-sampling Gaussian and regularity
conditions, the resulting estimator is $\sqrt{N}$-consistent and
asymptotically normal, with a nominal variance factor
captured by the single scalar $K(b)$.  At $b=1.5$, the AR(2) experiment
gives bias reductions of about $40\%$, $90\%$, and $96\%$ for the three
parameters at a nominal asymptotic variance cost of $16\%$
(Table~\ref{tab:kappa}).  The clean-data sample-size trends are
compatible with the theoretically predicted rate.  In the equal-budget
trimming experiment, oracle trimming is below robust estimation for weak and
moderate tones,
whereas robust estimation has lower aggregate RMSE for all plotted
$A\geq1.5$ without requiring knowledge of the disturbance frequencies.
The separate exact sampled-CAR(2) experiment provides a direct finite-sample
CT check; it does not replace the assumptions of the asymptotic theorem.

The Gumbel formulation opens several directions: non-stationary
contamination and joint model-order/\allowbreak robustness
selection (where the CT spectral roll-off makes the problem
especially demanding), multivariable CARMA with colored
excitation, and frequency-selective clipping that adapts~$b$
based on the local sensitivity $\|I^{-1}s_k\|$.
The approach assumes known model order ($n$, $m$).  The
gross-error contamination model suits persistent tonal disturbances;
bursty contamination may require time-varying extensions.
A reference implementation in Python, including the robust estimator,
$a(b)$ computation, and Monte Carlo scripts, is available at the
companion repository.

\appendix

\section{Technical Conditions}\label{app:lemmas}

The following conditions are assumed throughout.
The parameter space $\Theta$ is a compact subset of $\mathbb{R}^p$
with $\theta_0$ in its interior.
\begin{enumerate}
  \item \textit{(Smooth spectrum)} The applicable model spectrum
        $\Phi_k(\theta)$ is positive and $C^2$ in $\theta$ near
        $\theta_0$ at every retained frequency.  In the sampled setting it
        is the exact $f_\Delta$ in \eqref{eq:sampled-spectrum}.
  \item \textit{(Information)} $I(\theta_0)$ is positive definite,
        varies continuously in $\theta$, and converges to a bounded
        positive-definite limit as $N\to\infty$.
        Moreover, $\max_k\|s_k(\theta)\|^2 = O(\log N)$ uniformly
        near~$\theta_0$.  On the compact sampled frequency interval, a
        smooth exact sampled spectrum bounded away from zero gives the
        stronger $O(1)$ bound.
  \item \textit{(Periodogram asymptotics)} In the fixed-$\Delta t$
        long-span regime, the Gaussian stationary sampled process satisfies
        the frequency-domain uniform law and triangular-array CLT required
        below for the bounded score-weighted gradients.  In particular,
        every fixed collection of distinct retained positive-frequency
        normalised ordinates converges jointly to independent
        $\operatorname{Exp}(1)$ variables.  These are assumptions on the
        sampled process, not finite-sample identities; sampled state-space
        sufficient conditions are given in \cite{fasenMayer22}, with related
        frequency-domain conditions in \cite{dahlhaus96}.
  \item \textit{(Uniform criterion and dominated differentiation)}
        For the score used by the estimator, the family
        $\{\ell_q(\rho_k(\theta)): \theta\in\Theta\}$ (in particular
        $\ell_q=\ell_b$ for the proposed score) has an
        integrable envelope and is stochastically equicontinuous, so the
        criterion obeys the uniform law of large numbers used in
        Section~\ref{sec:robust}.  Differentiation under expectation is
        justified by local boundedness of $s_k$ and uniform integrability
        of $q_b(\rho_k)$ under $|q_b|\le b$.
  \item \textit{(Identifiability and canonical parameterisation)} Let
        $f_\theta$ denote the spectrum actually used for estimation, on its
        retained frequency domain $\mathcal D$.  The spectral map is
        injective in $L^2(\mathcal D)$: for any $\theta\ne\theta_0$ in the
        parameter space,
        \[
          \int_{\mathcal D} \bigl|\log f_\theta(\xi)
            - \log f_{\theta_0}(\xi)\bigr|^2\,d\xi > 0.
        \]
        For the CT spectral factor we restrict $A$ to be monic Hurwitz,
        $B$ to be monic Hurwitz (minimum phase), and $A,B$ to be coprime,
        with $\lambda>0$.  Absence of cancellations alone is insufficient: for
        example, the coprime numerators $s+1$ and $s-1$ have the same
        magnitude on the imaginary axis.  For sampled data we additionally
        assume injectivity of $\theta\mapsto f_\Delta(\cdot,\theta)$ and
        restrict CT state eigenvalues to the principal sampling strip
        $|\operatorname{Im}\lambda_i(A)|<\pi/\Delta t$; these restrictions
        exclude, rather than solve by assertion, phase and sampling aliases
        \cite{fasenMayer22}.
\end{enumerate}

\section{Proof of Theorem~\ref{thm:opt-huber}}\label{app:proof-opt-huber}

\begin{pf*}{Proof of Theorem~\ref{thm:opt-huber}}
Put $u:=\Lambda+\alpha_b$, $\psi_b:=\operatorname{clip}(u,-b,b)$,
$q:=u-\psi_b$, and abbreviate $c:=c_b$, $\eta_b:=\psi_b/c$ and
$B:=b/c$.  Centering and integration by parts under
$f_G'/f_G=-\Lambda$ give $E_G[\psi_b]=0$ and
$E_G[\psi_b\Lambda]=E_G[\psi_b']=c$, hence
$\eta_b\in\mathcal C_B$.

For part~(a), take any $\eta\in\mathcal C_B$ and set
$\delta:=\eta-\eta_b$.  The two normalisation constraints imply
$E_G[\delta u]=0$.  On the set where $q\ne0$, one has
$\eta_b=B\,\operatorname{sign}(q)$; since $|\eta|\leq B$, it follows
that $\delta q\leq0$ pointwise.  Thus
$E_G[\delta\eta_b]=-c^{-1}E_G[\delta q]\geq0$, and
$E_G[\eta^2]-E_G[\eta_b^2]
=E_G[\delta^2]+2E_G[\delta\eta_b]\geq0$.
Equality forces $\delta=0$ $G$-almost surely, proving uniqueness.

For part~(b), take any $\eta\in\mathcal C$ with
$M:=\|\eta\|_\infty<\infty$; the claim is immediate when $M=\infty$.
On the support of $q$, $\eta q\leq M|q|$ and
$\eta_bq=B|q|$, so $E_G[\delta q]\leq(M-B)E_G[|q|]$.
Equation~\eqref{eq:local-radius} and $b=cB$ give
$E_G[|q|]=r_b^2b=r_b^2cB$, hence
$E_G[\delta\eta_b]\geq-r_b^2B(M-B)$.  Substitution in the
variance decomposition yields
\[
  \mathcal R_{r_b}(\eta)-\mathcal R_{r_b}(\eta_b)
  \geq E_G[\delta^2]+r_b^2(M-B)^2\geq0.
\]
Again equality forces $\eta=\eta_b$ $G$-almost surely.  Finally,
$E_G[\eta_b^2]=E_G[\psi_b^2]/c_b^2=K(b)$ by
\eqref{eq:kappa}. \qquad$\square$
\end{pf*}

\begin{rmk}[Role of the symmetric clip for small $b$]
For $b\geq b_0$ the Gumbel score $e^z-1$ is bounded below by $-1$, so
the lower clip of $\psi_b$ is inactive on the support of
$\mathrm{Exp}(1)$.  For $0<b<b_0$ the score range
$|e^z - 1 + \alpha_b|$ exceeds $b$ in the lower tail, so the two-sided
clip is required.  The threshold $b_0\approx 0.797$ is the smallest $b$ for
which $1 - \alpha_b\leq b$; the practitioner-recommended range
$b\in[0.8,2.0]$ lies above it.  This distinction changes the geometry
of the optimal influence curve but not the normalisation or the proof
above.\end{rmk}

\section{Proof of Theorem~\ref{thm:asympvar}}\label{app:proof-asympvar}

\begin{pf*}{Proof of Theorem~\ref{thm:asympvar}}
\textbf{Lemma~C.1 (Central slope).}
\textit{For $\rho\sim\mathrm{Exp}(1)$ and the two-sided clipped Gumbel
score with thresholds $\rho_{-}\!:=\!\max(0,m_b{-}b)$ and
$\rho_{+}\!:=\!m_b{+}b$:}
\begin{align*}
  c(b) &:= \mathbb{E}[\rho\,q_b'(\rho)]
    = \int_{\rho_{-}}^{\rho_{+}} \rho\,e^{-\rho}\,d\rho \\
    &= (\rho_{-}{+}1)\,e^{-\rho_{-}} - (\rho_{+}{+}1)\,e^{-\rho_{+}} > 0.
\end{align*}
\textit{For $b\geq b_0$ this reduces to
$c(b) = 1 - (\rho_{+}{+}1)e^{-\rho_{+}}
      = 1 - (2{-}\alpha_b{+}b)\,e^{-(1-\alpha_b+b)}$,
the form used in the original (one-sided) closed-form derivation.}

For the general score in the theorem, put
$c_q:=E[\rho q'(\rho)]\ne0$ and
$S_T(\theta):=N^{-1}\!\sum_k s_k(\theta)q(\rho_k(\theta))$.
At $\theta_0$, $E[S_T]=0$ by the centering assumption.  Using
$\nabla_\theta \rho_k=-\rho_k s_k$ and noting that the term containing
$\partial s_k/\partial\theta^T$ has expectation zero by centering gives
\begin{equation*}
  E\!\left[\frac{\partial S_T}{\partial\theta^T}\Big|_{\theta_0}\right]
    = -c_q\,I_N(\theta_0),
  \qquad I_N=N^{-1}\!\sum_k s_ks_k^T.
\end{equation*}
Condition~4 and convergence of $I_N$ then yield
$\partial S_T/\partial\theta^T|_{\theta_0}
\xrightarrow{p}-c_q I(\theta_0)$.
Since $\hat\theta\xrightarrow{p}\theta_0$,
a mean-value expansion of $S_T(\hat\theta)=0$ about $\theta_0$ gives
\[
  0 = S_T(\theta_0)
    + \frac{\partial S_T}{\partial\theta}\Big|_{\tilde\theta}
      (\hat\theta-\theta_0),
\]
for some $\tilde\theta$ between $\hat\theta$ and $\theta_0$.
Because $\hat\theta\xrightarrow{p}\theta_0$ and $\partial S_T/\partial\theta$
is continuous in $\theta$ (Condition~1), we have
$\partial S_T/\partial\theta|_{\tilde\theta}
  \xrightarrow{p} -c_q\,I(\theta_0)$
by the uniform LLN (Condition~4).  Inverting and multiplying by
$\sqrt{N}$:
\begin{align*}
  \sqrt{N}\,(\hat\theta-\theta_0)
    &= \frac{I(\theta_0)^{-1}}{c_q}\cdot
    \frac{1}{\sqrt{N}}\sum_k s_k(\theta_0)\,
    q(\rho_k(\theta_0))\\
    &\quad + o_p(1).
\end{align*}
The bounded score and leverage condition give the Lindeberg control,
while the frequency-domain triangular-array CLT is assumed explicitly in
Condition~3; joint convergence of each fixed collection of ordinates alone
would not suffice.  That CLT gives a zero-mean normal limit with covariance
$E[q(\rho)^2]I(\theta_0)$ for the sum above.  Slutsky's theorem therefore
gives
\begin{align*}
  \sqrt{N}\,(\hat\theta-\theta_0)
    &\Rightarrow \mathcal{N}\!\Bigl(0,\,
    \tfrac{E[q(\rho)^2]}{c_q^2}\,I(\theta_0)^{-1}\Bigr) \\
    &= \mathcal{N}\!\bigl(0,K(q)\,I(\theta_0)^{-1}\bigr). \qquad\square
\end{align*}
\end{pf*}

\section{Efficiency Factor $K(b)$: Explicit Formula}\label{app:kappa}

For the two-sided clipped Gumbel score
$q_b(\rho) = \mathrm{clip}(\rho - m_b,\;-b,\;b)$ with
$m_b := 1 - \alpha_b$ and thresholds
$\rho_{-} := \max(0, m_b{-}b)$, $\rho_{+} := m_b{+}b$,
we reduce $K(b) = \mathbb{E}[q_b^2]/c(b)^2$ to a closed form on each
of the three regions $[0,\rho_{-})$, $[\rho_{-},\rho_{+}]$,
$(\rho_{+},\infty)$.  For $b\geq b_0$ the lower-clip region collapses
($\rho_{-}=0$) and the algebra below reduces to the one-sided closed
form (using $\alpha_b{+}\rho_{+} = 1{+}b$ and $e^{-\rho_{+}} = \alpha_b$
from the implicit branch of \eqref{eq:abrelation}); for $0<b<b_0$ the
lower-clip region contributes a third term but no new transcendentals
appear.

\noindent\textbf{Numerator $\mathbb{E}[q_b(\rho)^2]$.}
\begin{align*}
  \mathbb{E}[q_b(\rho)^2]
    &= b^2\bigl(1{-}e^{-\rho_{-}}\bigr)\mathbf 1_{\rho_{-}>0}
     + \!\!\int_{\rho_{-}}^{\rho_{+}}\!(\rho{-}m_b)^2 e^{-\rho}\,d\rho \\
    &\quad + b^2\,e^{-\rho_{+}}.
\end{align*}
The middle integral evaluates by integration by parts:
\[
  \!\!\int (\rho{-}m_b)^2 e^{-\rho} d\rho
   = -e^{-\rho}\!\bigl[(\rho{-}m_b)^2 + 2(\rho{-}m_b) + 2\bigr].
\]
For $b\geq b_0$ ($\rho_{-}=0$, $e^{-\rho_{+}}=\alpha_b$,
$\alpha_b{+}\rho_{+}=1{+}b$) this collapses to
$\mathbb{E}[q_b(\rho)^2] = (1-\alpha_b)^2 - 2\alpha_b b$,
recovering the one-sided closed form.

\noindent\textbf{Denominator $c(b)$.}
Since $q_b'(\rho)=1$ on $(\rho_{-},\rho_{+})$ and zero on the clipped
flats:
\begin{align*}
  c(b) &= \!\!\int_{\rho_{-}}^{\rho_{+}}\!\rho\,e^{-\rho}\,d\rho
        = (\rho_{-}{+}1)\,e^{-\rho_{-}} - (\rho_{+}{+}1)\,e^{-\rho_{+}}.
\end{align*}
For $b\geq b_0$ this reduces to $c(b) = (1-\alpha_b)^2 - \alpha_b b$.

\noindent\textbf{Result.}  $K(b) = \mathbb{E}[q_b(\rho)^2]/c(b)^2$.
For $b\geq b_0$ this gives the boxed one-sided form
\[
  K(b) = \frac{(1-\alpha_b)^2 - 2\alpha_b b}
              {\bigl[(1-\alpha_b)^2 - \alpha_b b\bigr]^2},
  \qquad \alpha_b\!=\!e^{-(1+b-\alpha_b)}.
\]
For $0<b<b_0$, substitute the closed-form
$\alpha_b = 1 - b - \log\!\bigl((1-e^{-2b})/b\bigr)$.  In both regimes
$K(b)$ is a finite combination of exponentials and polynomials in $b$.
Table~\ref{tab:kappa} lists the values for representative clipping
levels.

\section*{Declaration of generative AI and AI-assisted technologies in
the writing process}

During the preparation of this work the authors used Claude
(Anthropic) and Codex (OpenAI) to assist with the writing process,
mathematical consistency checks, and refactoring of the reference
Python implementation.  After using these tools, the authors reviewed
and edited the content as needed and take full responsibility for the
content of the publication.

\ack{The first author thanks Rik Pintelon (1959--2021) for hosting his
research stay at the Department ELEC, Vrije Universiteit Brussel;
Pintelon's work on frequency-domain system identification has been
foundational to the field.  Financial support from the Interuniversity
Attraction Poles programme, managed by the Belgian Federal Science
Policy Office, is gratefully acknowledged.}

\bibliographystyle{plain}
\bibliography{references}

\newpage

\biographyentry{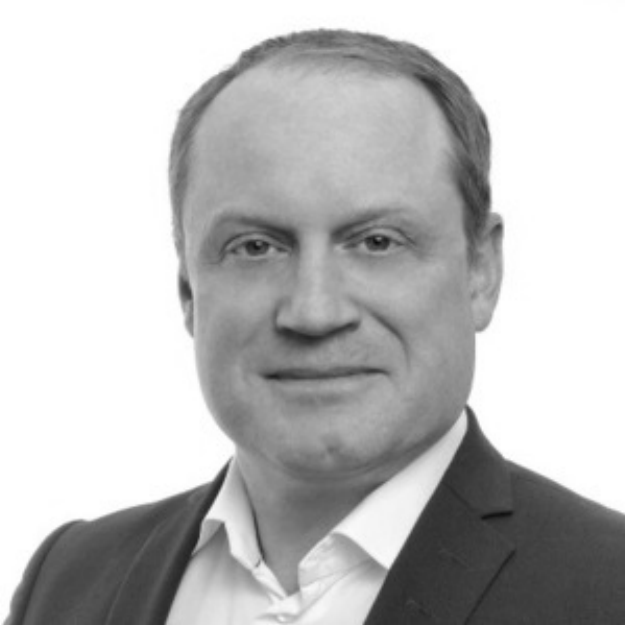}{Jonas Gillberg}{%
received the Ph.D.\ degree in Automatic Control from Link\"oping
University, Link\"oping, Sweden, in 2006, where his thesis work on
frequency-domain system identification was supervised by
Prof.~Lennart Ljung.  He was previously with IBM~Research in the
Quantum Computing division and is currently a Principal with ZyQE in
Stockholm, Sweden.  His research interests are
in system identification --- in particular robust estimation,
continuous-time methods, and frequency-domain techniques --- and in
convex optimization.}

\biographyentry{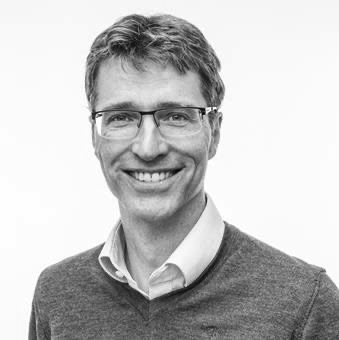}{Fredrik Gustafsson}{%
is professor in Sensor Informatics at the Department of Electrical
Engineering, Link\"oping University, since 2005.  He received the
M.Sc.\ degree in electrical engineering in 1988 and the Ph.D.\
degree in Automatic Control in 1992, both from Link\"oping
University.  During 1999--2005 he was professor in Communication
Systems.  He was awarded the Swedish Research Council Distinguished
Professor grant (r\aa dsprofessor) 2015--2024, was elected member of
the Royal Swedish Academy of Engineering Sciences (IVA) in 2007,
and became an IEEE Fellow in 2011.  His research interests are in
stochastic signal processing, adaptive filtering and change
detection, with applications to communication, vehicular, airborne,
audio, and wildlife systems.  He is the author of five books and
over 400 journal and conference papers, holds some 25 patents, and
co-founded NIRA Dynamics AB, Softube AB, Senion AB and the
tech-for-good initiative Ngulia.}

\end{document}